\documentclass[letterpaper,twocolumn,10pt]{article}
\usepackage{usenix2019_v3}
\usepackage[english]{babel}

\date{}
\usepackage{blindtext}

\usepackage{placeins}

\usepackage{xcolor}

\usepackage{booktabs} 

\usepackage{varwidth}
\usepackage{enumitem}
\usepackage{float}
\usepackage{graphicx}
\usepackage{rotating}
\usepackage[justification=centering]{caption}

\usepackage[font=footnotesize,format=plain,labelfont=bf,textfont=bf]{caption}
\usepackage[font=scriptsize]{subcaption}
\usepackage{amsfonts}
\usepackage{multirow}
\usepackage{listings}
\usepackage{mdwlist}
\usepackage{comment}
\usepackage[linesnumbered,ruled,vlined]{algorithm2e}
\usepackage{fancyvrb}
\SetKw{KwBy}{by}
\usepackage{relsize}
\usepackage{tabularx}
\newcolumntype{Y}{>{\centering\arraybackslash}X}
\usepackage{bold-extra}
\usepackage{tablefootnote}
\usepackage{hyperref}

\SetCommentSty{mycommfont}
\newcommand{\smartparagraph}[1]{\vspace{.05in}\noindent{\bf #1}}

\makecompactlist{itemize}{stditemize}
\usepackage{titling}
\usepackage{mdwlist}
\usepackage{multirow, makecell, colortbl}
\usepackage[most]{tcolorbox}

\tcbset{
    frame code={}
    center title,
    left=0pt,
    right=0pt,
    top=0pt,
    bottom=0pt,
    colback=gray!20,
    colframe=white,
    width=\dimexpr\columnwidth\relax,
    enlarge left by=0mm,
    boxsep=5pt,
    arc=0pt,outer arc=0pt,
    }
\usepackage{tikz}
\newcommand*\circled[1]{\tikz[baseline=(char.base)]{
            \node[shape=circle,draw,inner sep=2pt] (char) {#1};}}

\usepackage{cleveref}

\crefformat{section}{\S#2#1#3} 
\crefformat{subsection}{\S#2#1#3}
\crefformat{subsubsection}{\S#2#1#3}

\newcommand{\rl}{$\textsc{p4rl}$\xspace}
\newcommand{\lang}{$\texttt{p4q}$\xspace}
\newcommand{\system}{$\texttt{P6}$\xspace}
\newcommand{\agent}{$\textsl{Agent}$\xspace}
\newcommand{\reward}{$\textsl{Reward~System}$\xspace}

\newcommand{\fuzz}{{Fuzzer}\xspace}
\newcommand{\dete}{{Detector}\xspace}
\newcommand{\loc}{{Localizer}\xspace}
\newcommand{\tar}{{P4Tarantula}\xspace}
\newcommand{\pat}{{Patcher}\xspace}
\newcommand{\old}{{P$4_{14}$}\xspace}
\newcommand{\new}{{P$4_{16}$}\xspace}
\newcommand{\pd}{{platform-dependent}\xspace}
\newcommand{\pin}{{platform-independent}\xspace}

\begin{document}
\title{\Large \bf Towards Runtime Verification of Programmable Switches}
	\author{
		\rm Apoorv Shukla$^1$ \quad Kevin Hudemann$^{2,*}$ \quad Zsolt V\'agi$^{3,*}$ \quad Lily H\"ugerich$^1$ \quad Georgios Smaragdakis$^1$\\ \quad Stefan Schmid$^4$ \quad Artur Hecker$^5$ \quad Anja Feldmann$^{6,7}$ \\
		\footnotesize $^1$TU Berlin \quad $^2$SAP  \quad $^3$Swisscom \quad $^4$Faculty of Computer Science, University of Vienna \quad $^5$Huawei \quad $^6$MPI-Informatics \quad $^7$Saarland University\\
		\thanks {Kevin Hudemann and Zsolt V\'agi worked on this paper while they were affiliated with TU Berlin.}}
\maketitle
\begin{abstract}
\textit{Is it possible to 
patch software bugs in P4 programs without human involvement?} We show that this is partially possible in many cases due to advances in software testing and the structure of P4 programs. Our insight is that runtime verification can 
detect bugs, even those that are not detected at compile-time, with machine learning-guided fuzzing. 
This enables a more automated and real-time localization of bugs in P4 programs using software testing techniques like Tarantula. Once the bug in a P4 program is localized, the faulty code can be 
patched due to the programmable nature of P4. In addition, \pd bugs can be detected. 
From \old to \new (latest version), our observation is that as the programmable blocks increase, the patchability of P4 programs increases accordingly. To this end, we design, develop, and evaluate \system that 
(a) detects, (b) localizes, and (c) patches bugs in P4 programs with minimal human interaction. \system tests P4 switch non-intrusively, i.e., requires no modification to the P4 program for detecting and localizing bugs. We used a \system prototype to detect and patch seven existing bugs in eight publicly available P4 application programs deployed on two different switch platforms: behavioral model (bmv2) and Tofino. 
Our evaluation shows that \system significantly outperforms bug detection baselines while generating fewer packets and patches bugs in P4 programs such as \texttt{switch.p4} without triggering any regressions. 

%
\end{abstract}

\section{Introduction}
Programmable networks herald a paradigm shift in the design and operation of networks. The network devices on the data plane, e.g., switches, that traditionally have fixed and vendor-specific network functionality and rely on proprietary hardware and software, can now be programmed and customized by network operators. The P4 language~\cite{bosshart2014p4, p416specs} was introduced to enable the programmability and customization of data plane functionalities in network devices. P4 is an open-source domain-specific language designed to allow programming of packet forwarding planes, and is now supported by a number of network vendors. 

While programmable networks enable to break the tie between vendor-specific hardware and proprietary software, they facilitate an independent evolution of software and hardware. With the P4 language, one can define in a P4 program, the instructions for processing the packets, e.g., how the received packet should be read, manipulated, and forwarded by a network device, e.g., a switch. Nevertheless, with the new capabilities, new challenges in P4 software verification, i.e., ensuring that the software fully satisfies all the expected requirements, have been unleashed. The P4 switch behavior depends on the \emph{correctness} of the P4 programs running on them. 
We realize that a bug in a P4 program, i.e., a small fault such as a missing line of code or a fat finger error, or a vendor-specific implementation error,
can trigger unexpected and abnormal switch behavior. In the worst case, it can result in a network outage, or even a security compromise~\cite{peyman}. 

\noindent {\textit{\bf Problem Statement.}} In this paper, we pose the following question: \emph{``Is it possible to 
\texttt{detect}, \texttt{localize}, and \texttt{patch} software bugs in a P4 program without human involvement?''}. We believe that being able to answer this question, even partially, unlocks the full potential of programmable networks, improves their security, as well as increases their penetration in operational and mission-critical networks.

Recently, a panoply of P4 program verification tools~\cite{stoenescu2018debugging,liu2018p4v,neves2018verification,Neves:2018:VPP:3281411.3281421,Notzli:2018:PAT:3185467.3185497,p4consist} has been proposed. These verification systems, however, fail to 
repair the P4 program containing bugs.
Most of them~\cite{stoenescu2018debugging,liu2018p4v,neves2018verification,Neves:2018:VPP:3281411.3281421} 
aim to statically verify user-defined P4 programs which are later, compiled to run on a target switch. They mostly find bugs that violate the memory safety properties, e.g., invalid memory access, buffer overflow, etc. Furthermore, they are prone to false positives and are unable to verify the {\em runtime} behavior on real packets. In addition, classes of bugs, e.g., checksum-related, ECMP (Equal-Cost Multi-path) hash calculations-related or platform-dependent bugs, cannot be detected by static analysis approaches. 
Since, runtime verification aims to verify the \emph{actual} behavior against the \emph{expected} behavior of a switch by passing specially-crafted input packets to the switch and observing the behavior, 
such verification is complementary to static analysis. Note, the detection of bugs causing the abnormal runtime behavior is a complex and challenging task. In particular, the P4 switch does not throw any runtime exceptions. 
Furthermore, the detection of bugs can be a nightmare if there is no output, i.e., packets are silently dropped instead of being forwarded. Thus, the runtime verification of the switch behavior is crucial. 

A useful approach to verify the runtime behavior is fuzz testing or fuzzing~\cite{afl, lib, Godefroid:2012:SWF:2090147.2094081,peach,sulley,rad,zzuf,tfuzz,driller,shukla2019consistent, shukla2}, a well-known dynamic program testing technique that generates semi-valid, random inputs which may trigger abnormal program behavior. However, for fuzzing to be efficient, intelligence needs to be added to the input generation, so that the inputs are not rejected by the parser and it maximizes the chances of triggering bugs. This becomes crucial especially in networking, where the input space is huge, e.g., a 32-bit destination IPv4 address field in a packet header has $2^{32}$ possibilities. With the 5-tuple flows, the input space gets even more complex and large.
To make fuzzing more effective, we consider the use of machine learning, to guide 
the fuzzing process to generate smart inputs that trigger abnormal target behavior. Recently, Shukla et al.~\cite{shukla2} have shown that Reinforcement Learning (RL)~\cite{sutton2018reinforcement,Russell:2009:AIM:1671238} can be used to train a verification system. We build upon~\cite{shukla2} by adding (a) static analysis to the fuzzing process to significantly reduce the input search space, and thus, adding input structure awareness, and (b) support for \pd bug detection. 

Even if a bug in a P4 program is detected, the localization of code statements in the P4 code that are responsible for the bug, is non-trivial. The difficulty stems from the fact that practical P4 programs can be large with a dense conditional 
structure. In addition, the same faulty statements in a P4 program may be executed for both passed as well as failed test cases and thus, it gets hard to pinpoint the actual faulty line/s of code. 
Tarantula~\cite{jones2001visualization, jones2002visualization, jones2005empirical} is a dynamic program analysis technique that helps in fault localization by pinpointing the potential faulty lines of code. To localize the software bugs, we tailor Tarantula for generic software to P4 programs by building a localizer called \tar and integrating it with the bug detection of machine learning-guided fuzzing. 
In this paper, we combine these two approaches to detect and localize bugs in P4 programs in real-time. 

\smartparagraph{\system.} We, however, realize that automated program repair~\cite{goues2019automated} is an uncharted territory and becomes increasingly important as the software development lifecycle in programmable networks is short~\cite{dapipe} with insufficient testing. In this paper, we show that due to the structure of P4 programs, it is possible to automate the patching of \pin bugs in P4 programs, if the patch is available. 
To this end, we present \system, \emph{P4} with runtime \emph{P}rogram \emph{P}atching, a novel runtime P4-switch verification system that 
(a) \texttt{detects}, (b) \texttt{localizes}, and (c) \texttt{patches} software bugs in a P4 program with minimal human effort. \system improves the existing work based on machine learning-guided fuzzing~\cite{shukla2} 
in P4 by extending it and augmenting it with: (a) automated localization, and (b) runtime patching. \system relies on the combination of static analysis of the P4 program and Reinforcement Learning (RL) technique to guide the fuzzing process to verify the P4-switch behavior at runtime.

\smartparagraph{\system in a nutshell.} In \system, the first step is to capture the expected behavior of a P4 switch, which is accomplished using information from three different sources: (i) the control plane configuration, (ii) queries in \lang~(\cref{sec:lang}), a query language which we leverage to describe expected behavior using conditional statements, and (iii) accepted header layouts, e.g., IPv4, IPv6, etc, learned via static analysis of the P4 program.
If the \emph{actual} runtime behavior to the test packets generated via machine-learning guided fuzzing differs from the \emph{expected} behavior through the violation of the \lang queries, it signals a bug to \system which then identifies a patch from a library of patches. If the patch is available, \system modifies the original P4 program to no longer trigger the bug signaled by the \lang queries. Then, the patched P4 program is subjected to sanity and regression testing. 
We develop a prototype of \system and evaluate it by testing it on eight \new application programs from switch.p4~\cite{switch}, P4 tutorial solutions~\cite{p4tut}, and NetPaxos codebase~\cite{netp} across two P4 switch platforms, namely, behavioral model version~2~(bmv2)~\cite{bmv2} and Tofino~\cite{tofino}. Our results show that \system successfully detects, localizes and patches diverse bugs in all \new programs while significantly outperforming bug detection baselines without introducing any regressions. 


\smartparagraph{\bf Contributions.} Our main contributions are:

\noindent $\bullet$ We design, implement, and evaluate \system, the first end-to-end runtime P4 verification system that detects, localizes, and patches bugs in P4 programs non-intrusively. (\cref{sec:system})\\ 
$\bullet$ We observe that the success of \system relies on the increased patchability of P4 program from old (\old) to the new version (\new). 
We confirm this by studying the code of P4 applications in two different P4 versions (\old, \new). 
(\cref{sec:challenges})\\
$\bullet$ We present a \system prototype and report on an evaluation study. We evaluate our \system prototype 
on a P4 switch running eight \new programs (including \texttt{switch.p4} with $8715$ LOC) from publicly available sources~\cite{p4tut,netp,switch} across two platforms, namely, behavioral model and Tofino. Our results show that \system non-intrusively detects both \pd and \pin bugs, and significantly outperforms state-of-the-art bug detection baselines. (\cref{sec:evaluation})\\
$\bullet$ We show that in case of \pin bugs, \system can localize bugs and fix the P4 program, when the patch is available, without causing any regressions or introducing new bugs in an automated fashion. (\cref{sec:system}, \cref{sec:evaluation})\\
$\bullet$ To ensure reproducibility and facilitate follow-up work, we will release the \system software and library of ready patches for all existing bugs in the P4 programs to the research community. 

\begin{figure*}[ht]
    \centering
    \includegraphics[width=0.8\textwidth]{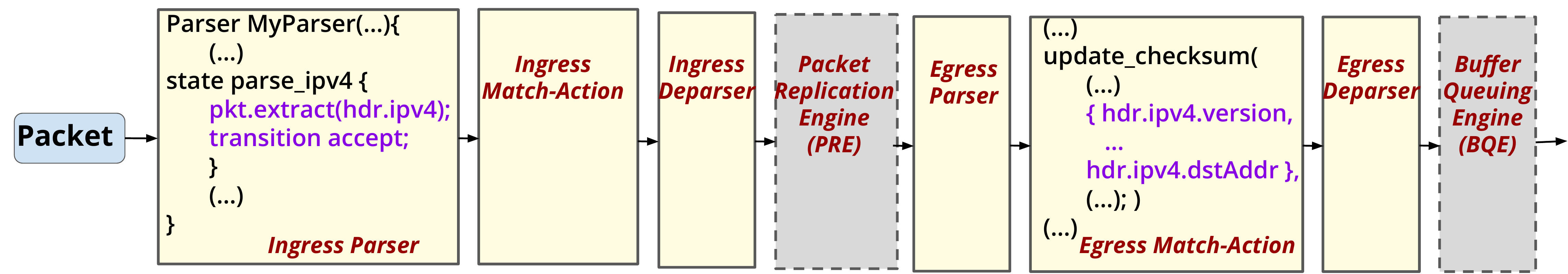}
    \caption{An example of a \pin 
    bug in \new packet processing pipeline.}
    \label{fig:mot2}
\end{figure*}

\begin{figure*}[ht]
    \centering
    \includegraphics[width=0.8\textwidth]{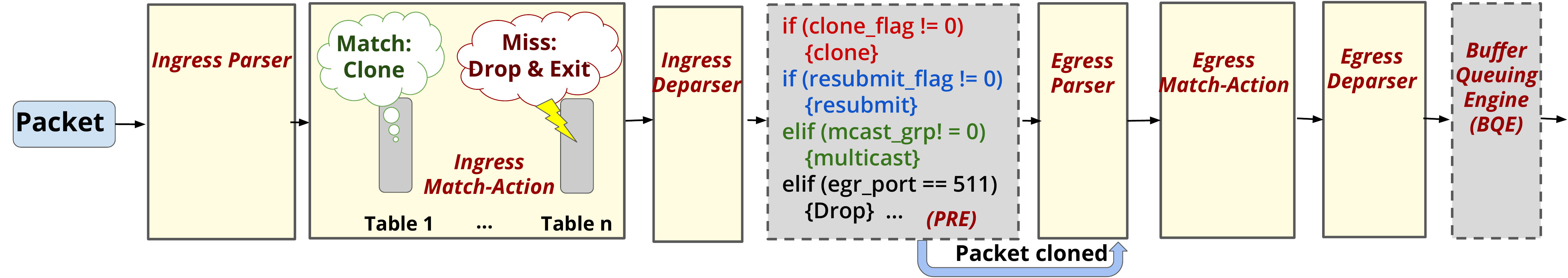}
    \caption{An example of a \pd 
    bug in \new packet processing pipeline.}
    \label{fig:mot3}
\end{figure*}

\section{Challenges \& Opportunities in P4 Verification}
\label{sec:challenges}
In this section, we describe the P4 packet processing pipeline. Then, we outline the challenges in discovering bugs in P4 programs or switches, and we motivate the need for runtime verification. We conclude this section by characterizing the evolution and structure of P4 programs, that, to our surprise provides opportunities for automated P4 program patching.
\subsection{Packet Processing Pipeline of P4}
P4~\cite{bosshart2014p4,p416specs} is a domain-specific language comprising of packet-processing abstractions, e.g., headers, parsers, tables, actions, and controls. The P4 packet processing pipeline evolved from~\cite{bosshart2013forwarding} to its current form \new~\cite{p416specs}, see Figure~\ref{fig:mot2}. In \new packet processing pipeline, there are six 
programmable blocks that are \pin, namely, {\tt ingress parser}, {\tt ingress match-action}, {\tt ingress deparser}, {\tt egress parser}, {\tt egress match-action}, and {\tt egress deparser}. The programmable blocks are annotated with a solid line in Figures~\ref{fig:mot2} and~\ref{fig:mot3}.

The ingress parser transforms the packet from bits into headers according to a parser specification provided by the programmer. After parsing, an ingress match-action (also called ingress control function) decides how the packet will be processed. Then, the packet is queued for egress processing in the ingress deparser. Upon dequeuing, the packet is processed by an egress match-action (also called egress control function). The egress deparser specification dictates how packets are deparsed from separate headers into a bit representation on output, and finally, the packet leaves the switch. Note that both ingress and egress match-actions (control functions) direct the packet through any number of match-action tables. 

In the \new packet processing pipeline, there are also two \pd blocks (annotated with dashed lines in Figures~\ref{fig:mot2} and~\ref{fig:mot3}), that rely on proprietary implementations of the hardware vendors and are non-programmable. These blocks are the {\tt packet re\-plication engine} (\texttt{PRE}) and the {\tt buffer queuing engine} (\texttt{BQE}).

\begin{figure*}[t]
\centering
\begin{subfigure}[l]{0.45\columnwidth}
\centering
\includegraphics[width=\columnwidth, height=4cm]{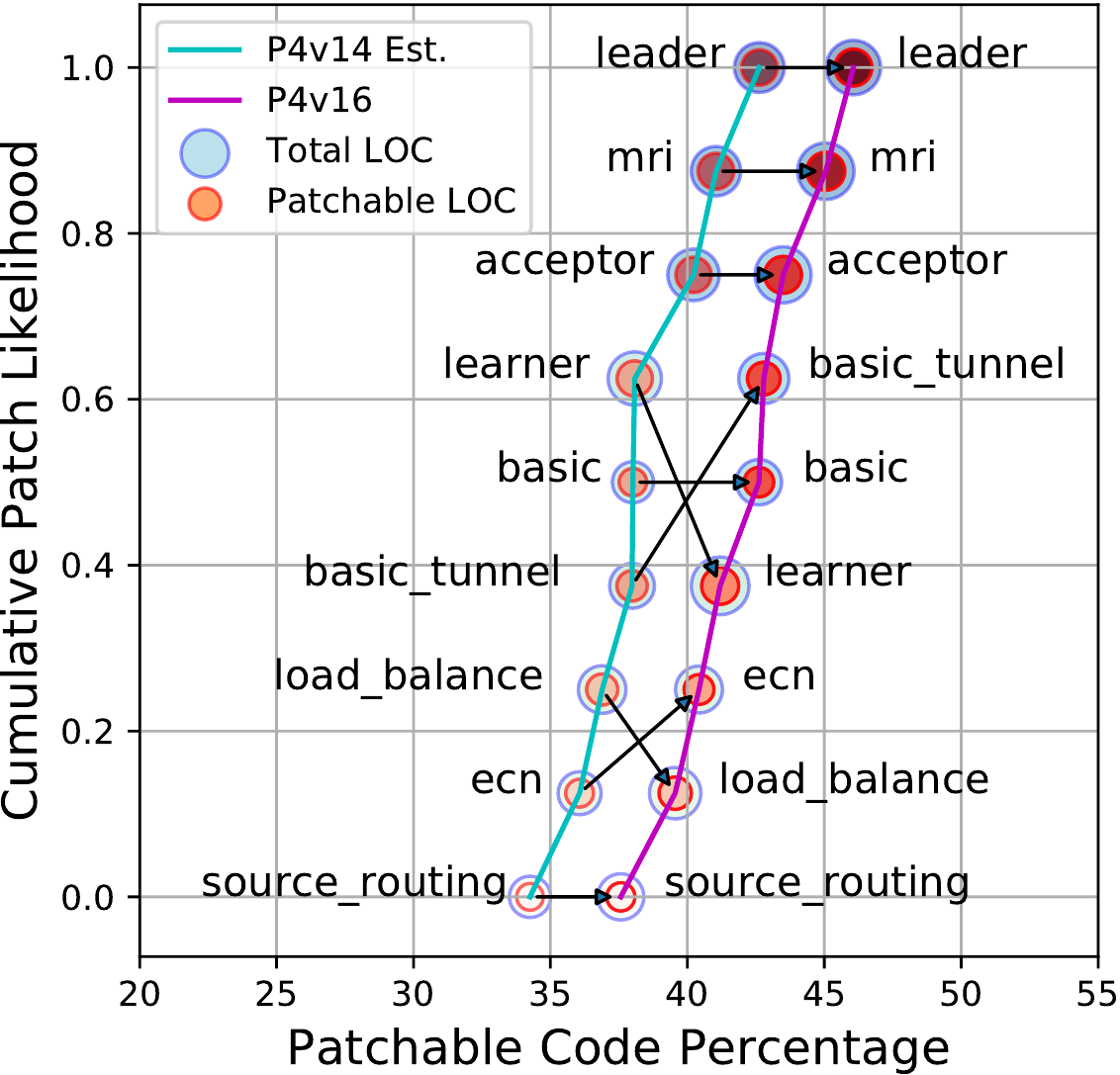}
\subcaption{The patchable code percentage increases for each P4 program from \old to \new in bmv2 switch platform~\cite{bmv2}, i.e., it does not include explicit ingress deparser and egress parser.}\label{fig:apps}
\end{subfigure}%
\hspace{0.01\textwidth}
\begin{subfigure}[c]{0.45\columnwidth}
\centering
\includegraphics[width=\columnwidth, height=4cm]{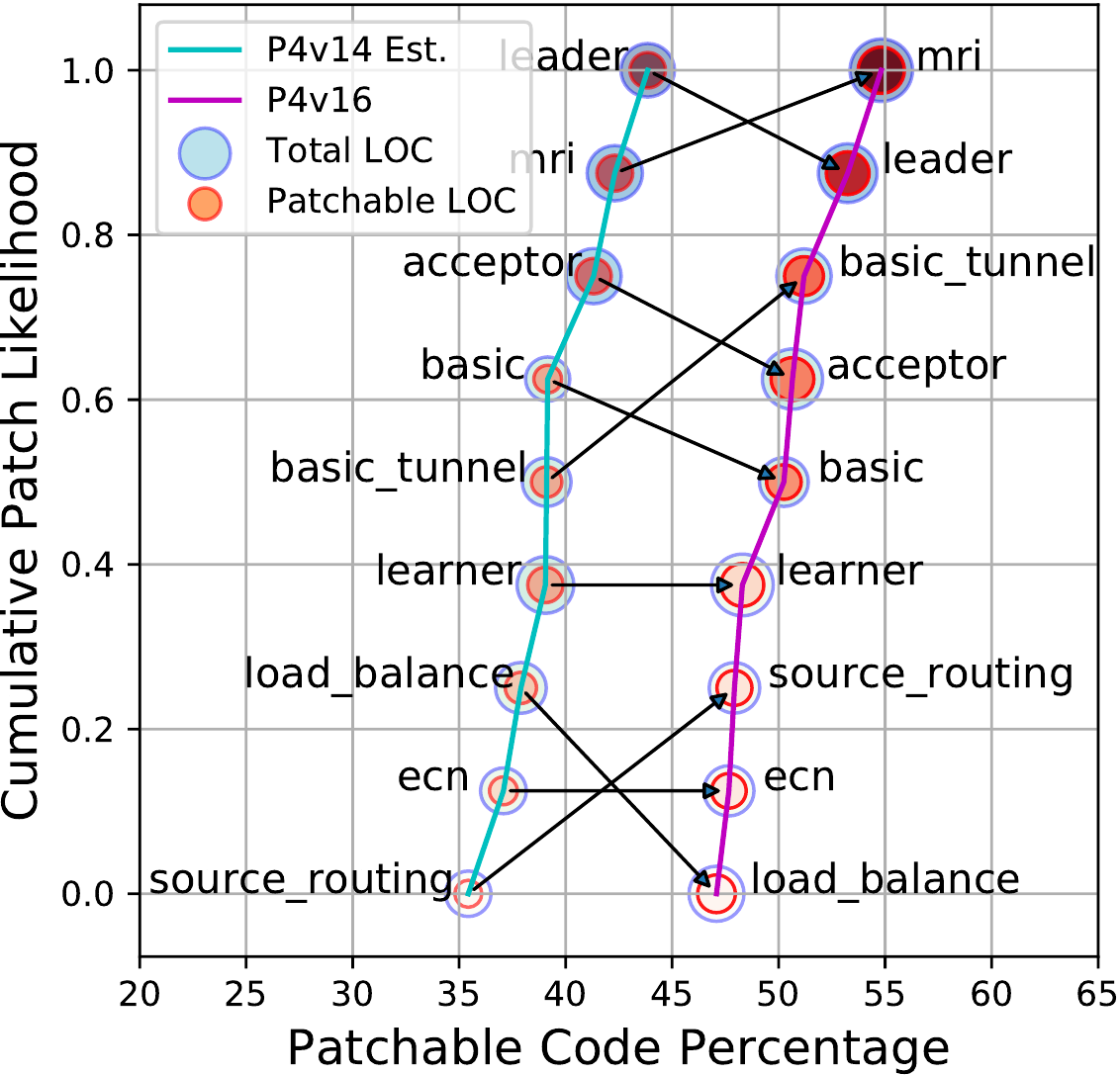}
\subcaption{The patchable code percentage increases for each P4 program from \old to \new in generic PSA switch platforms~\cite{psasec,psaing} like Tofino~\cite{tofino}, i.e., includes explicit ingress deparser and egress parser.}\label{fig:apps1}
\end{subfigure}%
\hspace{0.01\textwidth}
\begin{subfigure}[r]{0.47\columnwidth}
\centering
\includegraphics[width=\columnwidth, height=4cm]{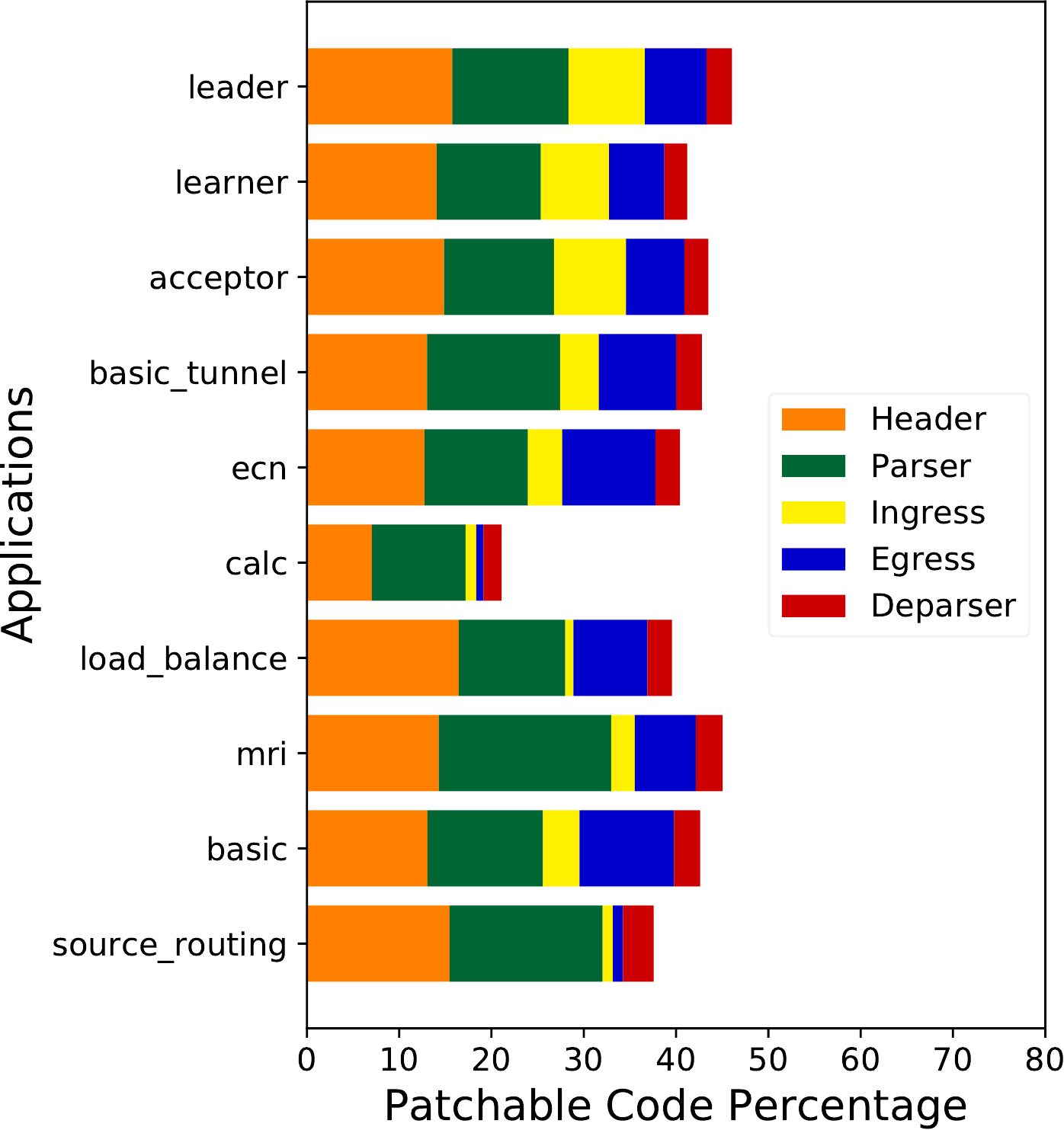}
\subcaption{The patchable code percentage for bmv2 switch platform~\cite{bmv2}, i.e., it does not include explicit ingress deparser and egress parser in all different \new programs from~\cite{p4tut,netp}.\label{fig:per}\newline }
\end{subfigure}
\hspace{0.01\textwidth}
\begin{subfigure}[r]{0.47\columnwidth}
\centering
\includegraphics[width=\columnwidth, height=4cm]{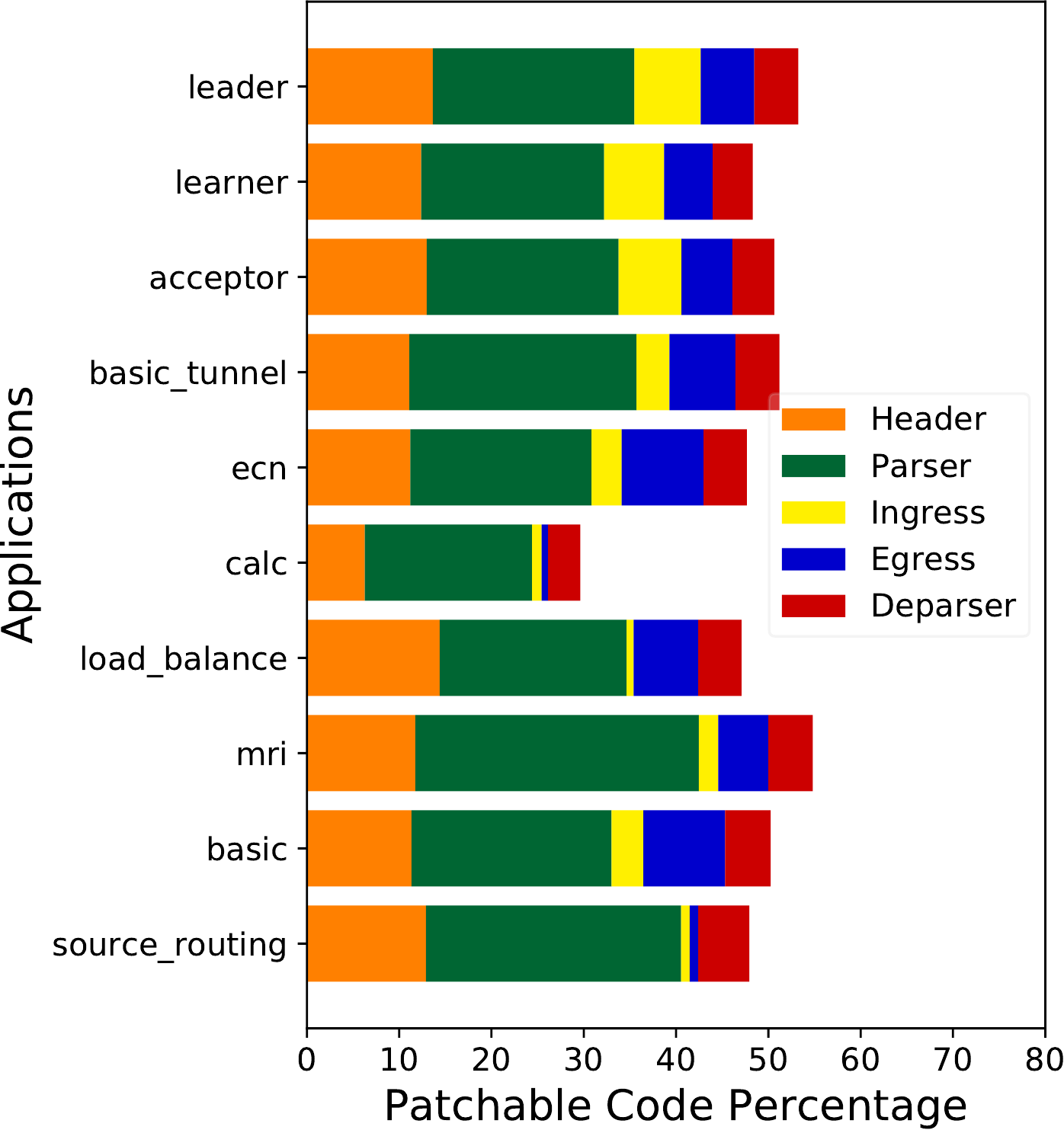}
\subcaption{The patchable code percentage for generic PSA switch platforms~\cite{psasec,psaing} like Tofino~\cite{tofino}, i.e., includes explicit ingress deparser and egress parser in all different \new programs from~\cite{p4tut,netp}. \label{fig:per1}}
\end{subfigure}
\caption{Evolution of the P4 program structure from \old to \new version.} 
\label{fig:struct}
\end{figure*}
\subsection{Challenges: Runtime Bugs in P4}
\label{sec:runtime_bugs}
 Bugs or errors can occur at any stage in the P4 pipeline. If a bug occurs in any of the programmable blocks, then the bug is \pin and software patching can solve the problem. If the bug appears in the non-programmable or \pd blocks, namely, the \texttt{PRE} or \texttt{BQE}, then the vendor has to be informed to fix the issue as the implementation is vendor-specific. 
P4 program verification systems~\cite{stoenescu2018debugging,liu2018p4v,neves2018verification,Neves:2018:VPP:3281411.3281421} are able to detect bugs using static analysis. Unfortunately, static analysis is (i) prone to false positives, (ii) cannot detect \pd bugs, and (iii) cannot detect runtime bugs that require to actively send real packets. 

\begin{figure}[h]
	\centering
	\includegraphics[width=0.65\columnwidth]{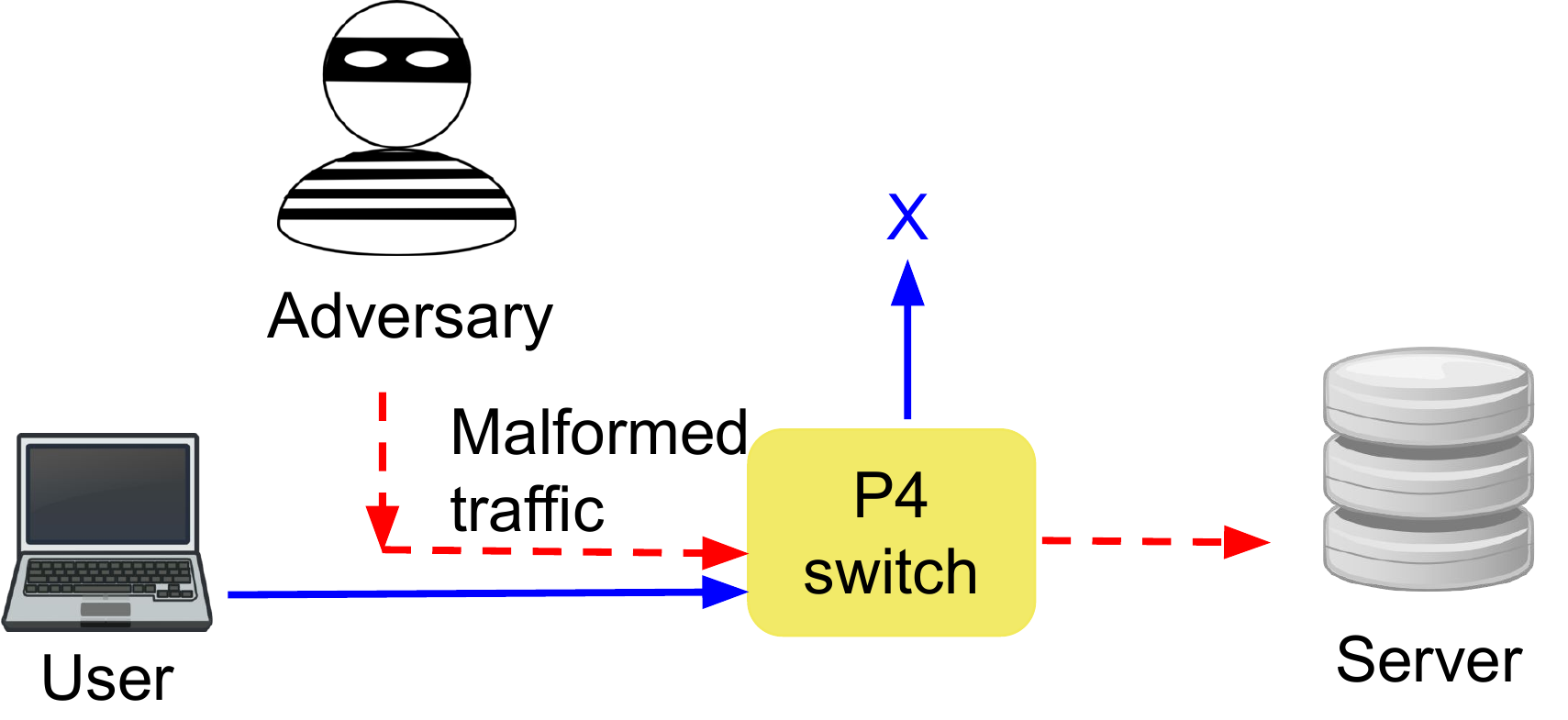}
	\caption{P4 switch running the P4 program does not check the faulty \texttt{IPv4 checksum} in the packet. Blue arrows show the expected path, red dashed arrows show the actual path.}
	\label{fig:example_checksum}
\end{figure}

As an example, consider the scenario in Figure~\ref{fig:mot2} (solid line blocks) that illustrates part of the implementation of Layer-3 (L3) switch, provided in the P4 tutorial solutions~\cite{p4tut}. Here, the parser 
does not check if the IPv4 header contains IPv4 options or not, i.e., if the IPv4 \texttt{ihl} field is equal to 5 or not. When updating the IPv4 \texttt{checksum} of the packets during egress processing, IPv4 options are not taken into account, hence for those IPv4 packets with options, the resulting \texttt{checksum} is wrong causing such packets to be forwarded and \textit{incorrectly} dropped at the next hop. This leads to anomalies in network behavior. Other bugs that fall in this category are those related to IPv4/6 \texttt{checksum} in the packet (see Figure~\ref{fig:example_checksum} later). 
Such bugs are \textit{\pin}, as they only result from programming errors. 


\begin{figure}[h]
	\includegraphics[width=\columnwidth]{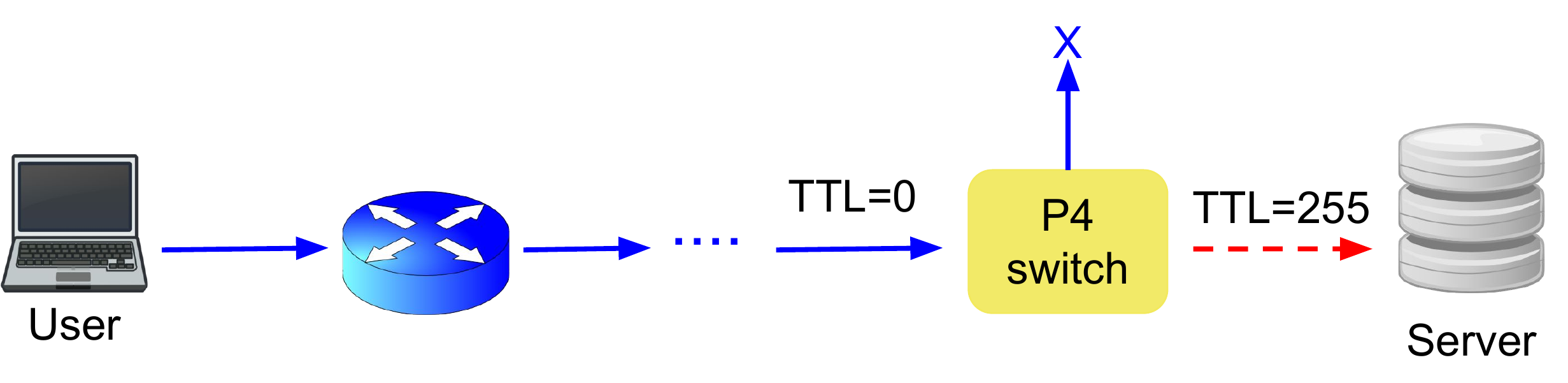}
	\caption{P4 switch running the P4 program does not check if the time-to-live (\texttt{TTL}) value in the packet is 0. Blue arrows show the expected, red dashed arrows show the actual path.}
	\label{fig:example_ttl}
\end{figure}

Figure~\ref{fig:example_ttl} illustrates a simple scenario where due to a bug or fault, the packet reaching a P4 switch has a time-to-live (\texttt{TTL}) field in the IP header with value as $0$. The expected behavior is that the packet gets dropped, but currently, there is no such check in the P4 code. Thus, the switch forwards the packet and decreases the value of the \texttt{TTL} field, causing it to be increased to $255$ incorrectly. This happens due to \texttt{wraparound} caused by underflow in an 8-bit field. Such an anomaly can be non-trivial to detect and localize. In addition, it can be responsible for the abnormal behavior of a P4 switch. 

As illustrated in Figure~\ref{fig:example_checksum}, the problem lies in the fact that the P4 program fails to specify that the \texttt{IPv4 checksum} inside the packet needs to be verified before forwarding. An adversary can easily intercept the packets and modify them (e.g., as a Man-in-the-Middle attack (MitM)
) and does not even need to recalculate the \texttt{checksum}. Therefore, additional information can be inserted even into encrypted packets. When such a malicious or malformed packet arrives at the P4 switch, it selects the corresponding action based on the match-action tables and forwards the packet without verifying the \texttt{checksum}. Such \texttt{checksum}-related bugs may inflict serious damages to critical servers and can be a nightmare to debug.

For a \textit{\pd bug}, consider the scenario shown in Figure~\ref{fig:mot3} (dashed line blocks). Here, we assume a P4 program implements at least two match-action tables. Any table except the last one could be a longest prefix match (\texttt{LPM}) table, offering unicast, clone and drop actions (ingress match-action block). The last match-action table implements an access control list (\texttt{ACL}). So, the packets can either be dropped or forwarded according to the chosen actions by the previous tables. In this case, it is possible that conflicting forwarding decisions are made. Consider packets are matched by the first table (Table 1) and a clone decision is made, later, those are dropped by the \texttt{ACL} table (Table n). In such a case, the forwarding behavior depends on the implementation of the \texttt{PRE}, which is \pd. The implementation of \texttt{PRE} of the SimpleSwitch target in the behavioral model (bmv2) is illustrated in Figure~\ref{fig:mot3}. It would drop the original packet, however, forward the cloned copy of the packet. Similar bugs can occur, if instead of the clone action, the resubmit action is chosen (blue). Similarly, another bug can be found when implementing multicast (green). 

The above motivates us to turn our attention to runtime detection of bugs. Runtime verification is a useful and complementary tool in the P4 verification repertoire that detects both \textit{\pin bugs} resulting from programming errors as well as \textit{\pd bugs}.

\subsection{Opportunities for Patching:\\
The Structure of a P4 Program}
In the evolution of P4, there are two recent versions: \old~\cite{p414specs} and \new~\cite{p416specs}. \new allows programmers to use definitions of a target switch-specific architecture, PSA (Portable Switch Architecture)~\cite{psasec,psaing}. 
\new is an upgraded version of \old. In particular, a large number of language features have been eliminated from \old and moved into libraries including counters, checksum units, meters, etc., in \new. \old allowed the programmer to explicitly program three blocks: ingress parser (including header definitions of accepted header layouts), ingress control and egress control functions. Recall that \new allows to explicitly program six programmable blocks (see Figure~\ref{fig:mot2}).
\begin{figure*}[t]
\centering
\includegraphics[width=0.85\textwidth]{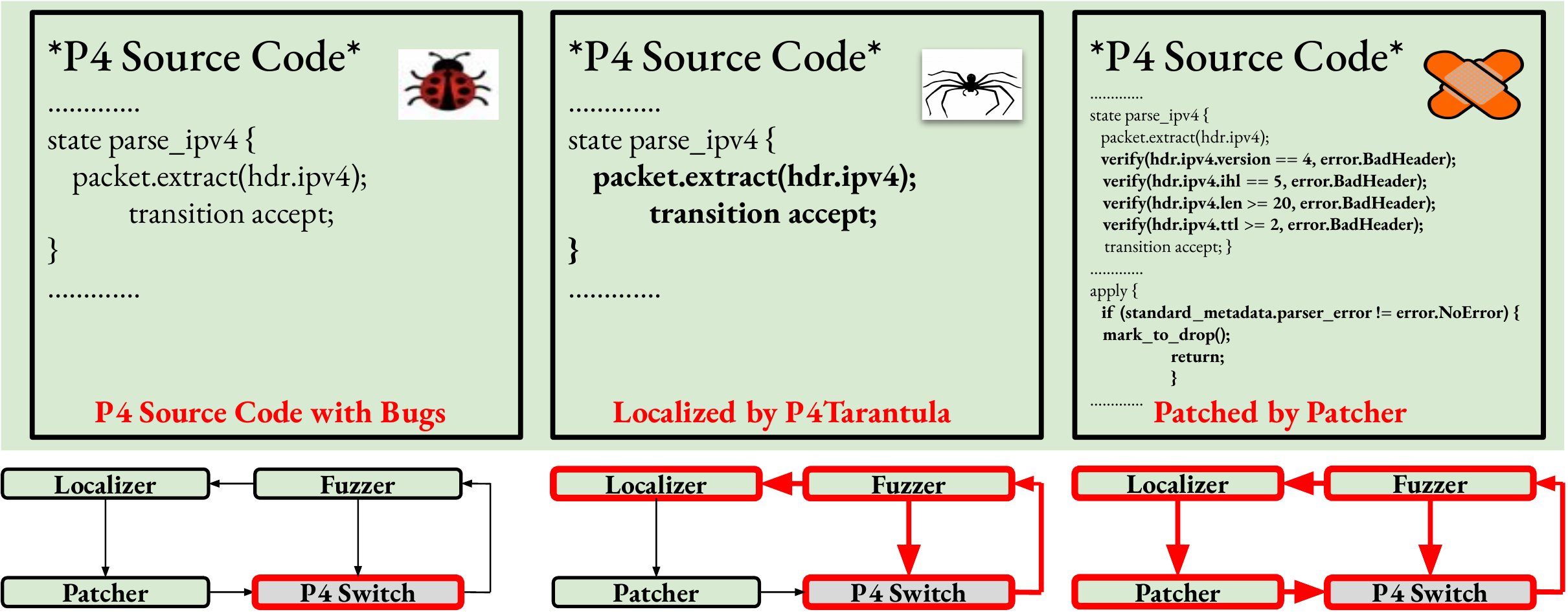}
\caption{\system in Action: depicting the automated detection, localization and patching of a bug in a L3 switch P4 program~\cite{p4tut}.}
\label{fig:work}
\end{figure*}
\begin{figure}[t]
\centering
\includegraphics[width=0.75\columnwidth]{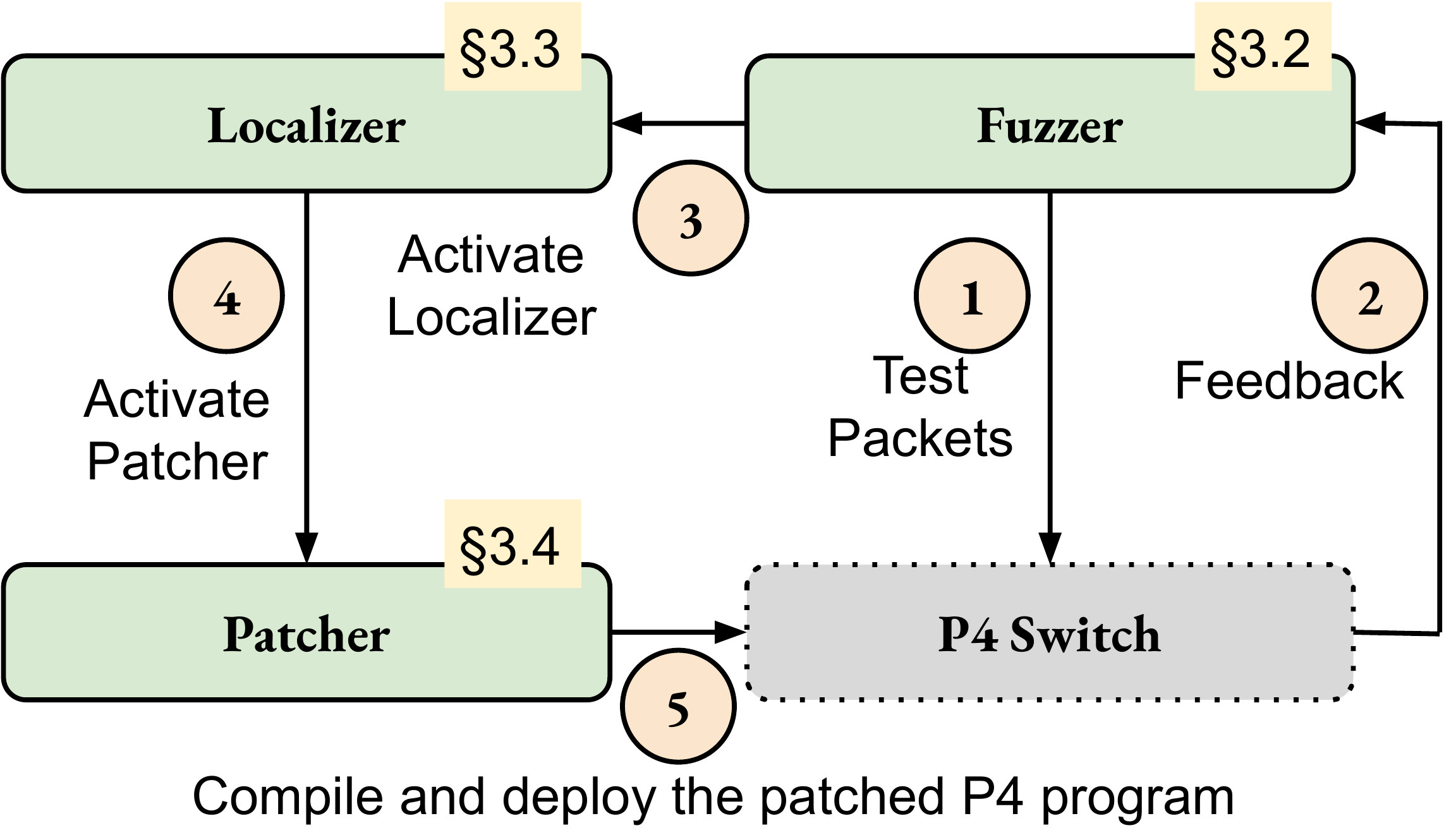}
\caption{\system Workflow. Modules of \system (in solid green boxes).}
\label{fig:sys}
\end{figure}

By analyzing programs in the \old and \new versions, we realize that as more blocks of the P4 program get programmable, there is more onus on the programmer to write a program that behaves as expected (when it gets compiled and deployed on the P4 switch). Missing checks or 
fat finger errors can cause havoc in the network. However, this is a blessing in disguise as the more programmable the code is, the more patchable it is. Thus, programming errors can be fixed. 
Figures~\ref{fig:apps} and~\ref{fig:apps1} illustrate that the potentially patchable code percentage increases from \old to \new in all applications (excluding calculator) from P4 tutorial solutions~\cite{p4tut} and NetPaxos codebase~\cite{netp} in behavioral model (bmv2) switch platform~\cite{bmv2} and other generic PSA switch platforms~\cite{psasec,psaing}, e.g., \sloppy{Tofino~\cite{tofino}} respectively. Figures~\ref{fig:per} and~\ref{fig:per1} illustrate the patchable code percentage in the latest \new version. The patchable code percentage comes from the six programmable blocks in \new. Roughly, whatever is programmable, is patchable. In principle, around 40-45\% of a P4 program is patchable in \new programs for behavioral model (bmv2) switch platform~\cite{bmv2} (Figure~\ref{fig:per}). This increases to 50-55\% if the ingress deparser and egress parser are programmable for other target switch platforms, e.g., Tofino~\cite{tofino} (Figure~\ref{fig:per1}). More importantly, the parser and header definitions account for 20-40\% of the total patchable code.\\
\emph{\textbf{Observation 1:} From} \old \emph{to} \new, the P4 \emph{program possesses twice as many programmable blocks doubling the opportunities for patching.}

If there is no bug in the parser/header code, the incoming packets with invalid header values will be dropped as expected and the packets with valid header values will be transmitted to the upcoming blocks else the invalid or semi-valid packets will incorrectly pass through the parser and may trigger abnormal runtime behavior. Assuming there was no bug in the parser/header code, those valid packets that are transmitted to the next blocks may still exhibit abnormal runtime behavior, if the programmable blocks containing, e.g., the application code logic, deparser have bug/s or due to bug/s in the platform dependent part which is vendor implementation-specific. \\ 
\emph{\textbf{Observation 2:} Once, a bug is detected and localized in the \pin part of a P4 program, it is patchable; a \pd bug is not patchable, however, the vendor can be informed if and when detected.}

\section{\system: System Design}\label{sec:system}
\subsection{\system: Overview}\label{sec:overview}
The goal of \system (see Figure~\ref{fig:sys}) is to detect, localize and patch the software bugs in a P4 program at runtime with minimal human effort. This is achieved by verifying the \emph{actual} runtime behavior against the \emph{expected} behavior of a P4 switch running a pre-compiled P4 program to the incoming packets. 

 The \system system contains three main modules:\\
\textbf{(1) \fuzz}: Generates test packets using RL-guided fuzzing, static analysis, and \lang queries~(\Cref{sec:lang}) to the P4 switch running the pre-compiled P4 program. (\Cref{sec:fuzzer})\\
\textbf{(2) \loc}: \tar is the \loc which pinpoints faulty lines of code causing bugs in the P4 program. (\Cref{sec:tarantula})\\
\textbf{(3) \pat}: Automates patching of the bugs localized by \tar~\loc, if patchable. Then, \pat compiles and loads the patched P4 program on the P4 switch. (\Cref{sec:patcher})

\smartparagraph{\system Workflow.} \system is a closed-loop control system. Through a pre-generated dictionary from control plane configuration, \lang queries, and static analysis of a P4 program, the expected runtime behavior of the P4 switch is captured and sent as an input to the \fuzz containing the RL \agent and the \reward (\Cref{sec:fuzzer}). As shown in Figure~\ref{fig:sys}, the \fuzz selects appropriate mutation actions such as add/delete/modify bytes in a packet to generate test packets towards the P4 switch running the pre-compiled P4 program \circled{1}. If the actual runtime behavior towards the packets defies the expected behavior through the violation of the \lang queries, it signals a bug in the form of a reward as a feedback to the \reward which is then, exploited by the RL \agent to improve the training process by selecting better mutation actions on the packet \circled{2}. After the bug detection, the \fuzz automatically triggers \loc (\Cref{sec:tarantula}), \tar (only for \pin bugs; for \pd bugs, the vendor is informed) which pinpoints the faulty line of code \circled{3} to trigger the \pat (\Cref{sec:patcher}) which searches for the appropriate patch from a library of patches for the corresponding P4 program \circled{4}. If the patch is available, \pat modifies the original P4 program, compiles and loads it on the P4 switch and checks if the bug is no longer triggered by \lang queries by repeating the whole-cycle and executing sanity and regression testing \circled{5}. 
Note, \system is non-intrusive and thus, requires no modification to the P4 program for testing before patching. 

\smartparagraph{\system in Action.} Before we dive into the details of \fuzz, \loc and \pat, we demonstrate the operation of \system. 
Figure \ref{fig:work} illustrates how \system detects, localizes, and patches an existing bug in a layer-3 (L3) switch P4 source code (program) from~\cite{p4tut} in an automated fashion. The left part of Figure \ref{fig:work} shows the P4 program containing a \pin bug in the parser code, i.e., no header field validation is implemented, hence all IPv4 packets are \emph{incorrectly} accepted by the parser. After the P4 program is deployed on the P4 switch, \system is triggered. Initially, the \fuzz detects the bug violating the corresponding \lang query based on the feedback (reward) received from the P4 switch. Then, it triggers the \tar for localization (shown in the center of Figure \ref{fig:work}) where it pinpoints the problematic part of the code (highlighted). Afterwards, the \pat is triggered automatically, patching the necessary problematic parts of the code, i.e., adding header field verification statements (highlighted in right), after checking if the patch was indeed missing from the P4 program. Finally, \pat automatically compiles~\cite{p4c} and deploys the patched P4 program on the P4 switch, and triggers \system to ensure that the patches caused no regressions and fixed the detected bug.
\subsection{\fuzz: RL-guided Fuzzing}
\label{sec:fuzzer}
The goal of \fuzz is to detect the runtime bugs discussed in~\Cref{sec:runtime_bugs}. We improve \cite{shukla2} by augmenting \fuzz with the static analysis of a P4 program which makes the \fuzz aware of the input structure or accepted header layouts, e.g., IPv4, IPv6, etc. and thus, it significantly reduces the input search space. Indeed, techniques to further reduce the input search space within the accepted headers are discussed in~\cite{alembic}, which can be augmented to static analysis.
We guide the mutation-based white-box fuzzing~\cite{Godefroid:2012:SWF:2090147.2094081} via RL~\cite{sutton2018reinforcement,Russell:2009:AIM:1671238}. The feedback in the form of \emph{rewards} is received from the switch based on the evaluation of \emph{actual} against \emph{expected} runtime behavior. Note, the expected behavior is determined using the static analysis, the control plane configuration, i.e., forwarding rules and \lang queries (\Cref{sec:lang}). \lang queries are conditional queries (\texttt{if-then-else}) where each query has multiple conditions and each condition acts as a test case. A violation of a test case represents a bug detection. 

\smartparagraph{Reinforcement Learning (RL).} 
Reinforcement learning~\cite{sutton2018reinforcement,Russell:2009:AIM:1671238} is a machine learning technique that aims at enabling an \agent to learn how to interact with an environment, based on a series of reinforcements, i.e., rewards or punishments received from the target environment, in our case, a switch. The \agent observes the switch and chooses an action to be executed. After the action is executed, the \agent receives a reward or punishment from the switch. 
While the goal of learning is to maximize the rewards, we argue it is equally crucial to design a machine learning model which is general enough for any kind of target environment. To detect the bugs triggered by fuzzing, one can observe the output of the target switch in response to the input packets. Thus, reinforcement learning allows developing a \reward where feedback in the form of rewards from the switch trains the \agent and thus, guides the fuzzing process.

In our RL-based model, we define states, actions, and rewards as follows:
\begin{figure}[t]
\centering
\includegraphics[width=0.85\columnwidth]{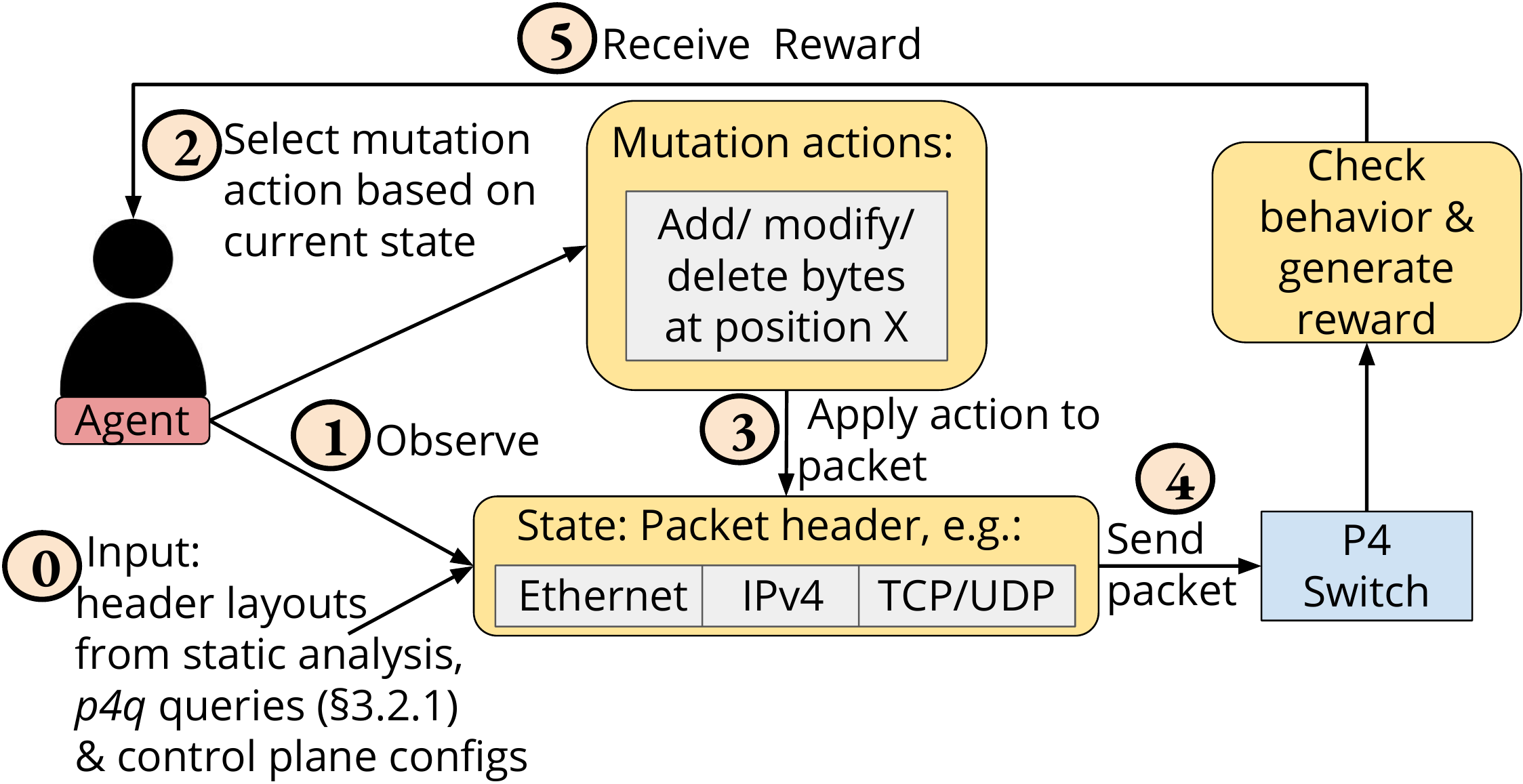}
\caption{\fuzz. \reward (in yellow) and \agent (in pink).}
\label{fig:fuzz}
\end{figure}

\smartparagraph{States:} The sequence of bytes forming the packet header. 

\smartparagraph{Actions:} The set of mutation actions for each individual packet header field, e.g., add, modify or delete bytes at a given position in the packet header. Note, the add and modify actions either use random bytes or bytes from a pre-generated dictionary (explained below).



\smartparagraph{Rewards:} The \agent can immediately receive the reward, after a mutated packet was sent to the target switch and the results of the execution are evaluated. It is likely to experience sparse rewards when most of the sent packets do not trigger any bug. Thus, the reward is defined as $0$, if the packet did not trigger a bug and $1$, if the packet successfully triggered a bug. 

The input to the \fuzz is a dictionary (hereafter, referred to as \texttt{dict}) that comprises information extracted from static analysis, the control plane configuration, and the queries defined with p4q~(\Cref{sec:lang}). 
The static analysis is used to derive the input structure awareness such as accepted header layouts and available header fields in the P4 program.
The control plane configuration comprises the forwarding table contents and the \pd configuration. Boundary values for the header fields may be extracted from the \lang queries, i.e., when queries explicitly compare packet header fields with values, e.g., \texttt{TTL} $>$ $0$.


Figure \ref{fig:fuzz} depicts the \fuzz workflow. In step 0 (initialization), the \reward receives the \texttt{dict} as an input. Then, the \agent observes the current state or the current packet header (see the initialization in \cref{sec:agent_alg}). The observed state is the input for the neural networks of the \agent (\Cref{sec:agent_alg}), which outputs the appropriate mutation action. The selected action is applied for the given packet, and the packet is sent to the P4 switch. After the packet is processed by the switch, the behavior is evaluated, the reward of $1$ is generated when the \lang query specifying the expected behavior is violated and returned to the \agent. In particular, the packet which was sent to the P4 switch is saved together with a final \emph{verdict} (pass or fail). A packet's \emph{verdict} is considered \textit{either} passed: if the generated reward is equal to $0$, i.e., \emph{actual} runtime behavior matches \emph{expected} behavior when the \lang query is not violated \textit{or} failed: if the generated reward is equal to $1$, i.e., \emph{actual} runtime behavior does not match \emph{expected} behavior when the \lang query is violated. Then, the \agent (\Cref{sec:agent_alg}) uses the received reward to improve the action selection in subsequent executions (exploitation).

\begin{figure}[t]
\centering
\includegraphics[width=0.62\columnwidth]{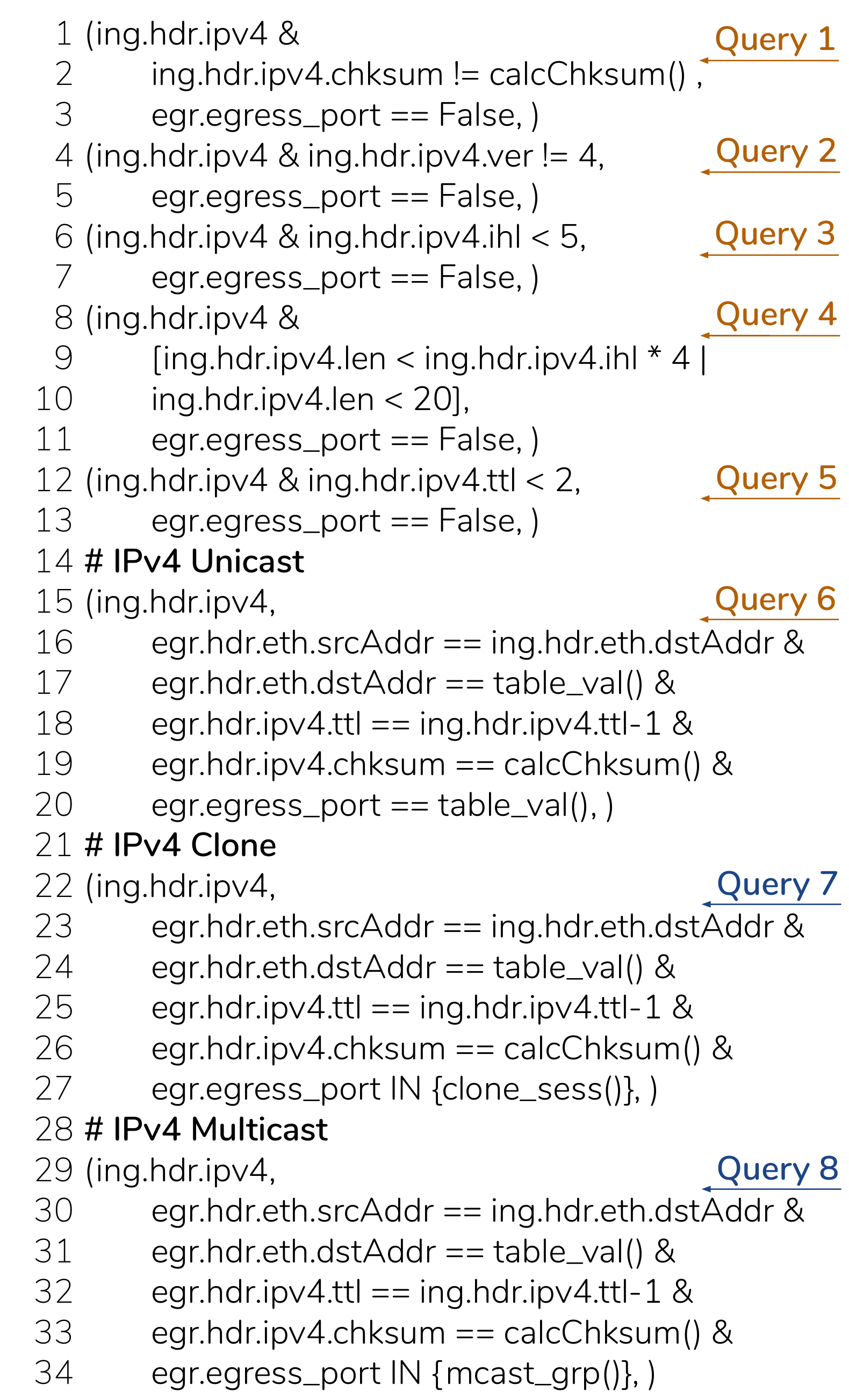}
\caption{\lang Queries. Queries 1-6 represent \pin, and Query 7-8 represent \pd queries respectively.}
\label{fig:langf}
\end{figure}
\subsubsection{\lang: Query Language}
\label{sec:lang}
Before diving into the details of the \agent training, we explain the query language, \lang~\cite{shukla2}, used for specifying the \textit{expected} switch behavior. To achieve the goal of an automated runtime verification system, \system system must query the \emph{actual} runtime behavior of a P4 switch against a specification defining the \emph{expected} behavior. To extend the query repertoire of \lang from~\cite{shukla2}, we augment it with \pd queries. In a nutshell, \lang queries are used to compare \textit{expected} against \textit{actual} switch behavior. 
 
\smartparagraph{\lang queries.} In a \lang query, the behavior is expressed using \texttt{if-then-else} statements in the form of tuples. The programmer can specify conditions for packets to fulfill at ingress of the switch (\texttt{if}), with corresponding conditions to fulfill at egress (\texttt{then}). In addition, the programmer can describe alternative conditions (\texttt{else}), e.g., if the condition of the \texttt{then} branch is not fulfilled at egress. \emph{To automate the usage of} \system, \emph{an option to execute all the queries of} \lang \emph{with a single command is provided} (see \Cref{sec:prototype}). 
To define these conditions, the \lang syntax and grammar are used.

\begin{figure}[t]
\centering
\includegraphics[width=\linewidth]{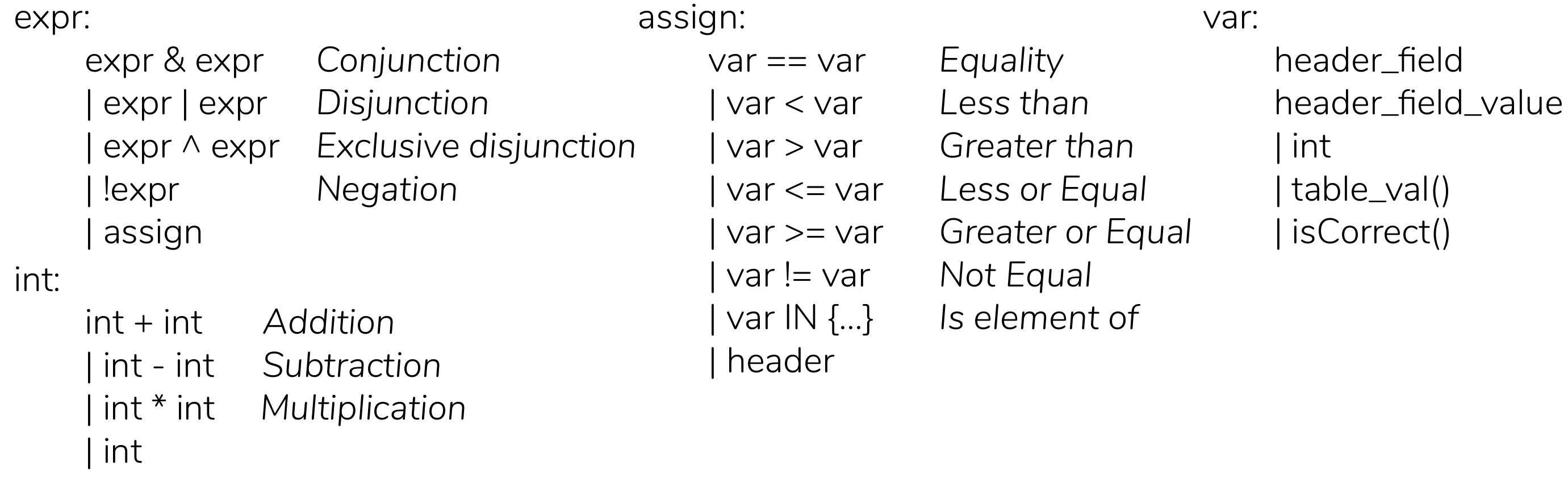}
\caption{\lang Grammar.}
\label{fig:langg}
\end{figure}

\smartparagraph{\lang Grammar.}
Figure~\ref{fig:langg} depicts the grammar and constructs defined in \lang. The \lang grammar allows common boolean expressions and relational operators as they can be found in many programming languages like C, Java or Python, to ease the work for the programmer. The boolean expressions and relational operators have the same semantics as common logical operators and expressions.
Variables can either be integers, header fields, header field values, or the evaluation result of the primitive methods, e.g., \texttt{calcChksum()} and \texttt{table\_val()}. Each header has a prefix (ing. or egr.) indicating if it is the packet arriving at ingress or exiting the switch at egress.\\


Figure~\ref{fig:langf} illustrates an example of how the packet processing behavior of an IPv4 layer 3 (L3) switch, written in P4, can be queried easily using \lang. Query 1 (lines 1-3), defines that incoming packets with a wrong IPv4 \texttt{checksum} are expected to be dropped.
Similarly, the following four queries (lines 4-13) express the validation of the IPv4 version field, the IPv4 header length, the packet length and the IPv4 time-to-live (\texttt{TTL}) field for packets at ingress of the switch respectively. However, there are also conditions for packets at the egress of the switch. These conditions are described by Query 6 (lines 15-20). Namely, changing source and destination \texttt{MAC} addresses to the correct values, decrementing the \texttt{TTL} value by 1, recalculating the IPv4 \texttt{checksum} and emitting the packet on the correct port as instructed by the control plane configuration (forwarding rules). Query 7 (lines 22-27) corresponds to the \pd part of the switch (\texttt{PRE}) and defines conditions for packets that are cloned by the switch. Such packets need to fulfill the same conditions as per Query 6, but the egress port should correspond to the clone session configuration of the target switch. Similarly, Query 8 (lines 29-34) expresses the conditions for multicast packets that need to fulfill the same conditions as per Query 7 but the egress ports should correspond to the configured multicast group configuration of the target switch.

Note, queries written in \lang can be extended, reused and provided in the form of libraries. 
More importantly, the \lang queries help in the availability of a library of pre-defined patches for the corresponding violations. Note, an easily extensible interface is provided to augment \lang further with user queries as per the deployment scenario to allow detection of more bugs.
\begin{algorithm}[t]
\scriptsize
\DontPrintSemicolon
\caption{\agent Training}
 \label{alg:algo_1}
 \KwIn{Empty prioritized experience replay memory $M$, uninitialized online and target network}
 \KwOut{Trained online and target network models}
 Initialize online network with random weights\;
 Initialize target network with copy of online network parameters\;
 \For{$i=1$ \KwTo $num\_episodes$}{
  Initialize byte sequence $b_{1}$\;
  Preprocess $b_{1}$ to get the initial state $s_{1} = preprocess(b_{1})$\;
  \For{$step = 1$ \KwTo $max\_ep\_len$}{
   Select action $a_{step}$ randomly with probability $\epsilon$ (exploration) or use online network to predict $a_{step}$ (exploitation)\;
   Execute action $a_{step}$, observe reward $r_{step}$ and byte sequence $b_{step+1}$\;
   Set $s_{step+1} = preprocess(b_{step+1})$\;
   Save the transition $(s_{step}, a_{step}, r_{step}, s_{step+1}, terminal)$ in $M$\;
   Sample batch of transitions $(s_{j}, a_{j}, r_{j}, s_{j+1})$ from $M$\;
   $y_{j} =              
   \begin{cases} r_{j} \quad \text{if terminal} \\
    r_{j} + \gamma * Q(s_{j+1}, \max_{a} Q(s_{j+1},a_{j}; \Theta), \Theta') \quad \text{otherwise}
   \end{cases}
   $\;
   Perform stochastic gradient descent using categorical cross entropy loss function\;
   }
 }
\end{algorithm}
\subsubsection{\agent}
The \agent houses the RL algorithm (Algorithm~\ref{alg:algo_1}), 
which is inspired by Double Deep Q Network (Double DQN)~\cite{Hasselt:2016:DRL:3016100.3016191}, an improved version of Deep Q Networks (DQN)~\cite{mnih2013playing}.\\
\smartparagraph{Double Deep Q Network (DDQN).}
DDQN algorithm~\cite{Hasselt:2016:DRL:3016100.3016191} is a recently-developed algorithm based on Q-learning~\cite{Russell:2009:AIM:1671238}, hence a \textit{model-free} reinforcement learning algorithm. \textit{Model-free} means, the \agent does not need to learn a model of the dynamics of an environment and how different actions affect it. This is beneficial, as it can be difficult to retrieve accurate models of the environment. At the same time, the goal is to provide sample efficient learning, i.e. reduce the number of packets sent to the target switch, makes the DDQN a suitable choice.
The basic concept of the algorithm is to use the current state (packet header) as an input to a neural network, which predicts the action the \agent shall select to maximize future rewards. In addition, Double DQN algorithm splits action selection in a certain state from the evaluation of that action. To achieve, it uses two neural networks: (i) the online network responsible for action selection, and (ii) the target network evaluating the selected action. This improves the learning process of the \agent, as overoptimism 
of the future reward when selecting a certain action, is reduced and 
thus, helps to avoid overfitting. 

\smartparagraph{Prioritized Experience Replay.} Experience replay~\cite{lin1993reinforcement} was introduced to eliminate problems of oscillation or divergence of parameters, resulting from correlated data. To overcome this problem, the experiences of the \agent, i.e., a tuple comprising the current state, predicted action, reward received, and resulting state are saved in the memory of \agent.
To enable learning by experience replay, the neural network model is updated using random samples from past experiences.
To counter the scenario of sparse rewards, a simple form of prioritized experience replay, inspired by~\cite{schaul2015prioritized}, is applied. The memory is sorted by absolute reward and each experience is prioritized by a configurable factor and the index. 

\smartparagraph{\agent Training Algorithm.}\label{sec:agent_alg}
Algorithm~\ref{alg:algo_1} presents the training algorithm of the \agent in the \system system. For our algorithm, we rely on the use of Multi-Layer Perceptron (MLP)~\cite{Rumelhart:1986:LIR:104279.104293}. In the initialization phase, the weights of online and target neural networks are initialized (Lines 1-2). For each execution, the current state is reinitialized by randomly choosing a packet header in byte representation from a pre-generated set of packet headers. The corresponding bytes are then converted to a sequence of float representations (Lines 4-5). An $\epsilon$-greedy policy is applied to determine the action to be executed (Line 7). Applying an $\epsilon$-greedy policy means that during training of the \agent, an action is selected randomly by the \agent with probability $\epsilon$ to ensure sufficient exploration. As the training progresses, probability $\epsilon$ is decreased linearly until a lower bound is reached. This helps in reducing overfitting as well, since the \agent never stops exploring the effects of other actions on the environment during training. The determined action will be executed, the result is observed and saved in the experience memory (Lines 8-10). As a last step, a sample out of the experience memory is selected to calculate $y_{j}$ which is used to calculate the categorical cross-entropy loss and perform the stochastic gradient descent step to update the network weights (Lines 11-13). 

\begin{algorithm}[t]
\scriptsize
	\SetAlgoLined
	\KwIn{P4 source code ($SC$), sent packets ($Ps$) and corresponding \emph{verdicts} ($V$)}
	\KwOut{S[j] - suspiciousness score for the corresponding line $j$}
	\tcp{V[p] represents the \emph{verdict} about packet p (pass or fail)\\
		SC[j] represents line j of the source code\\ 
	\textbf{Initialization}}
	$totalFailed = 0, totalPassed = 0$\\
	\ForEach{p in Ps}{
	\eIf{V[p] == pass}{$totalPassed += 1$}{$totalFailed += 1$}
	follow $p$ through $SC$:\\
	\ForEach{executed line j in SC}{
		\eIf{V[p] == pass}{$SC[j].pass += 1$}{$SC[j].fail += 1$}
	$S[j] = \frac{SC[j].fail / totalFailed}{SC[j].pass / totalPassed + SC[j].fail / totalFailed}$}}
	call $\pat$
	\caption{\tar (\loc)}
	\label{algorithm:tarantula}
\end{algorithm}
\subsection{\loc: \tar}
\label{sec:tarantula}
\tar is the \loc or the bug localization module of \system. \tar is based on a dynamic program analysis technique for generic software, Tarantula~\cite{jones2001visualization,jones2005empirical}. In case a bug is discovered by \fuzz, it automatically notifies \tar. \emph{Note, \tar will not be notified in case of \pd bugs, as they are neither localizable nor patchable}.  
As an input, \tar uses the P4 program or source code, the packets that were sent by \fuzz as per the \lang query (test cases) to trigger the bug and the pass or fail \emph{verdict} corresponding to those sent packets. Recall, a \emph{verdict} corresponds to a condition of the \lang query which acts as a test case. 

Algorithm \ref{algorithm:tarantula} presents the localization algorithm used by \tar. First, \tar initializes two counters, measuring the number of passed or failed \emph{verdicts} corresponding to the sent packets (Line 1). In the next step, \tar increments the counters according to the \emph{verdicts} made for the given packet (Lines 3-7). Now, the P4 source code needs to be traversed line-by-line (similar to symbolic execution but with actual packet header values to avoid all possible header values), to find the code execution path for the given packet (Line 8). 
For each line in the P4 source code that is executed for the given packet, counters for the corresponding \emph{verdicts} are incremented (Lines 10-14). For the executed lines of the P4 source code, a \textit{suspiciousness score}~\cite{jones2005empirical} is calculated (Line 15). The suspiciousness score is between $0$ and $1$ as the same line/s can be executed for passed and failed \emph{verdicts} corresponding to packets. This score corresponds to the likelihood that a line of code is causing a potential bug. The closer it is to $1$, the more likely it is that the corresponding line of code is problematic. 
Finally, the P4 source code lines are ordered as per their suspiciousness score to localize the bug. Then, \pat is notified.\\ 
\smartparagraph{More details on code traversal by \tar.}
While implementing \tar, we accurately traverse the P4 program execution
path for any given packet. To overcome, we have implemented our code-traversal solution as a part of a 
script responsible for calculating the suspiciousness scores for the lines of the source code of the P4 program. 
We follow the execution
in the source code of the P4 program from the ``start'' state of the parser until the
end of the execution, i.e., the deparser stage of the P4 packet processing pipeline.
By following the execution of the program as soon as it receives a given packet containing header values, we can determine how the different conditions in a P4 program are evaluated for the packet and follow the correct branch of the P4 program at different branching points.

\subsection{\pat}
\label{sec:patcher}

\pat is the novel automated patching module of the \system system. 
If a bug is localized by \tar, it 
notifies \pat. The input for \pat is the P4 source code, the results of static analysis of the P4 source code, the localization results of \tar, and the violated \lang query. 
\pat compares the localized problematic parts of the code with appropriate available patches. Note, \pat comes with a library of patches for P4 programs, i.e., those which violate \lang queries. Nevertheless, it can be easily extended when, previously unseen bugs, e.g., bugs in application code logic, are detected. 

From the results of the static analysis, \pat can extract the needed parser state names, header names, header field names, metadata names and metadata field names for the patches in the current version of the library of patches. In P4, metadata is used to pass information from one of the programmable or non-programmable blocks to another. 

Note, in most P4 programs (including the publicly available programs from~\cite{p4tut,netp, switch}) no variables apart from user-defined names for parser states or header/metadata fields are present. Thus, with the gathered knowledge about user-defined names \pat can compare through, e.g., regex or string comparison, 
if the patch (correct code) is already present in the P4 source code or if missing, the patch needs to be applied. Note, if the patches in the patch library require the analysis of custom variables or stateful components, e.g., registers and meters, the comparison if the patch is present or not requires further analysis of the code. 

In case no appropriate patch is available, the programmer is informed by the \pat. After \pat finishes the execution, it calls the P4 compiler (\texttt{p4c}) to re-compile the patched version of the P4 program and triggers the re-deployment of the code on the P4 switch. In addition, the \fuzz is notified 
by \pat to test the patched program again, to confirm the patches and ensure no regressions were caused by the patches by testing via the \lang queries and executing regression testing. 

A patch has the following properties: (a) preferably, few lines of code, e.g., missing checks in parser, (b) makes the P4 program conform to the expected behavior, (c) passes the sanity testing or checks for basic functionality, and (d) does not cause regressions breaking existing functionality elsewhere.

\begin{algorithm}[t]
	\scriptsize
	\SetAlgoLined
	\KwIn{P4 source code ($SC$), static analysis results ($Sr$), localization results ($Lr$) and violated \lang query ($q$)}
	\KwOut{A patched version of the source-code ($PSC$)}
	\tcp{The patcher 
	offers a patch only for those lines where the suspiciousness score $\geq 0.5$}
	Import \& process user-defined parser state names, header and header field names, metadata and metadata field names from $Sr$ required for patches in the patch-library\\
	\For{lines in $Lr$}{
	\eIf{Suspiciousness score $\geq 0.5$}{check corresponding line/s of code pinpointed by $\tar$\\
	\eIf{the patch is missing and violating $q$}{apply the preferred patch}{inform the programmer}
	Goto next line}{Goto next line}
	}	Compile \& re-deploy the patched P4 program ($PSC$) and notify $\fuzz$ for testing the patches and regressions
	\caption{\pat}
	\label{algorithm:patcher}
\end{algorithm}

Algorithm \ref{algorithm:patcher} shows the \pat algorithm. First, \pat imports the needed header or metadata field names, as well as parser state names for the currently available patches in the library of patches. Then, for each line in the localization results, \pat checks if the suspiciousness score is greater than the defined threshold of $0.5$\footnote{Threshold is configurable as per deployment scenario.} (Line 2), as it is highly likely that the corresponding line of code is responsible for triggering the detected bug. 
In case the suspiciousness score is above the defined threshold, \pat will check the corresponding line of code. The \pat, then, checks if the patch is available, 
e.g., through string comparison with the appropriate patch to be applied for the violated \lang query. If the patch is indeed missing, then the problematic line of code is patched, else the programmer is informed as the appropriate patch is not available (Lines 3-8). 
Once, all the localization results are processed (Lines 9-12), the patched P4 program is 
compiled by triggering the compiler (\texttt{p4c}) to be re-deployed on the P4 switch and the \fuzz is triggered to re-test the patched code (Line 14).

\smartparagraph{When is the \pat automated?} Currently, the library of patches exists for \pin bugs, i.e., those violating \lang queries 1-6 in Figure~\ref{fig:langf}. 
If the bug exists in a deployment-specific application code logic, then the programmer can be informed by the \pat to provide patches. Recall, for \pd bugs, the \tar will not be invoked and the vendor will be informed accordingly.

\section{\system Prototype}\label{sec:prototype}
We develop a \system prototype using Python version 3.6 with $\approx3,100$ lines of code (LOC); \fuzz with $\approx2,200$ LOC, \tar with $\approx490$ LOC and \pat with $\approx430$ LOC. \fuzz is implemented using Keras~\cite{keras} library with Tensorflow~\cite{tensor} backend and Scapy~\cite{scapy} for packet generation and monitoring. Currently, \system only supports programs written in \new~\cite{p416specs} as the P4 compiler (p4c \cite{p4c}) supports the translation of programs written in \old~\cite{p414specs} to \new. The \agent was trained separately, for each condition of each query written in \lang. The training process as well as the later execution using the trained Agents, however, can be parallelized.
For queries described in Figure~\ref{fig:langf}, the trained model of the \agent can be reused for testing different P4 programs that implement IPv4 packet processing.

In addition to the modules described in~\cref{sec:system}, we implement a control plane module using P4Runtime~\cite{p4runtime} and Python. For the P4 switch, we rely on software switches supporting \new, namely behavioral model (bmv2)~\cite{bmv2} with SimpleSwitchGrpc target (Version 1.12.0), and Barefoot Tofino Model~\cite{tofino} (Version 8.3.0).
To simplify and automate the usage of \system, a default option is provided where all queries of \lang 
are executed by:~\texttt{\system `p4/source\_code\_location' --default}
\begin{table}
\centering
\scriptsize
\begin{tabular}{|*{5}{c|}}
\hline
Bug IDs & Bugs & Queries (Figure~\ref{fig:langf})\\
\hline
1 & \tt{Accepted wrong checksum} (PI)  &  Query 1 \\
\hline
2 & \tt{Generated wrong checksum} (PI) & Query 6 (Line 19) \\
\hline
3 & \tt{Incorrect IP version} (PI) & Query 2 \\
\hline
4 & \tt{IP IHL value out of bounds} (PI) & Query 3 \\
\hline
5 & \tt{IP TotalLen value is too small} (PI) & Query 4 \\
\hline
6 & \tt{TTL 0 or 1 is accepted} (PI) & Query 5 \\
\hline
7 & \tt{TTL not decremented} (PI) & Query 6 (Line 18) \\
\hline
8 & \tt{Clone not dropped} (PD)  & Query 7 (Line 27) \\
\hline
9 & \tt{Resubmitted packet not dropped} (PD)  &  Query 6 (Line 20) \\
\hline
10 & \tt{Multicast packet not dropped} (PD)  & Query 8 (Line 34) \\
\hline
    \end{tabular}
    \caption{Bugs (with Bug IDs) detected by the \system prototype through the violation of the corresponding \lang queries (in Figure~\ref{fig:langf}). Note, PI and PD refer to \pin and -dependent respectively.} 
    \label{tab:packet_types}
\end{table}

\section{Evaluation}\label{sec:evaluation}
In this section, we evaluate the verification capabilities of \system.
\subsection{Baselines}\label{sec:baselines}
We compare \system against three baseline fuzzing approaches: 

\smartparagraph{(1) Advanced Agent.} The first baseline is an Advanced Agent only relying on random fuzz action selection, i.e., without prioritized experience replay. Thus, Advanced Agent can 
execute the same mutation actions as \system, 
but cannot learn which actions lead to rewards. It represents the intelligent baseline.

\smartparagraph{(2) IPv4-based fuzzer.} The second baseline is an IPv4-based fuzzer, which is aware of the IPv4 header layout and randomizes the different available header fields, except IP options fields and the destination IP as it prevents the packets from being dropped by the forwarding rules of the P4 switch. The actual behavior is evaluated using the queries of 
\lang.  

\smartparagraph{(3) Na\"ive fuzzer.} The third baseline is a simple na\"ive fuzzer, which is not aware of any packet header layouts. It generates and sends Ethernet frames from purely random mutation of bytes. The actual behavior is evaluated using the \lang queries. 
\subsection{Bugs} 
Table~\ref{tab:packet_types} provides an overview of \emph{existing} bug types (with bug IDs) detected in the publicly available P4 programs from~\cite{switch,p4tut, netp} by the \system prototype. These bugs are detected as they violate the corresponding \lang queries (from Figure~\ref{fig:langf}). 
In total, \system prototype can detect $10$ distinct bugs in the P4 programs. 
Out of these $10$ bugs, $7$ are patchable \pin (bugs $1-7$), and $3$ are \pd bugs (bug $8-10$). 

\smartparagraph{Platform-independent bugs.}
The two detected bugs with bug ID $1$ and $2$, are related to wrong IPv4 \texttt{checksum} computation and missing \texttt{checksum} validation. \system is able to detect, localize and patch these bugs. 
The four bugs with IDs $3-6$ are missing or wrong IPv4 packet header validation. Specifically, missing validation of IP version (bug $3$), IPv4 header length (bug $4$), IPv4 total length (bug $5$) and IPv4 time-to-live (\texttt{TTL}) (bug $6$). 
\system can detect, localize and patch these bugs. While the current approaches \emph{may} be able to detect the bugs, they still lack localization and patching of the bugs. The last of the \pin bugs is faulty \texttt{TTL} decrement bug (bug $7$).
In case, the P4 program accepts packets with IPv4 \texttt{TTL} $0$, the \texttt{TTL} decrement is still executed, causing an incorrect increment of \texttt{TTL} to $255$. Note, these $7$ \pin bugs already exist in the publicly available P4 programs of~\cite{p4tut, netp}. In \texttt{switch.p4} program~\cite{switch}, bugs with IDs 1,2,4 and 5 exist.

\smartparagraph{Platform-dependent bugs.}
In addition to the aforementioned \pin bugs, \system is able to detect three \pd bugs.
The first bug (Bug ID $8$) is described in Figure~\ref{fig:mot3} and occurs when using ingress-to-egress clone action. It violates Query $7$ in Figure~\ref{fig:langf} leading to incorrect forwarding of cloned packets when they are supposed to be dropped. The second bug (Bug ID $9$) involves the resubmit operation. Packets with the resubmit metadata field set which are marked to be dropped in a later stage, will be incorrectly resubmitted, i.e., the packet will be processed again, starting at the ingress parser. Thus, packets that are not expected to be resubmitted will be processed again. In the worst case, this can lead to packets being resubmitted over and over again or other unexpected behavior. It violates Query $6$ in Figure~\ref{fig:langf}. The third bug (Bug ID $10$) 
involves the multicast operation. If a packet is marked to be dropped and later the \textit{mcast\_grp} metadata field is set, then the multicasted copies of the packet are incorrectly forwarded and do not get dropped. It violates Query $8$ in Figure~\ref{fig:langf}. Note, we found all \pd bugs specifically, in \texttt{basic.p4} program~\cite{p4tut} on bmv2 platform~\cite{bmv2}. Furthermore, these bugs \textit{cannot} be detected by current approaches that are based on static analysis.

\begin{figure*}[t]
\centering
\begin{subfigure}[c]{0.66\columnwidth}
\centering
\includegraphics[width=\columnwidth]{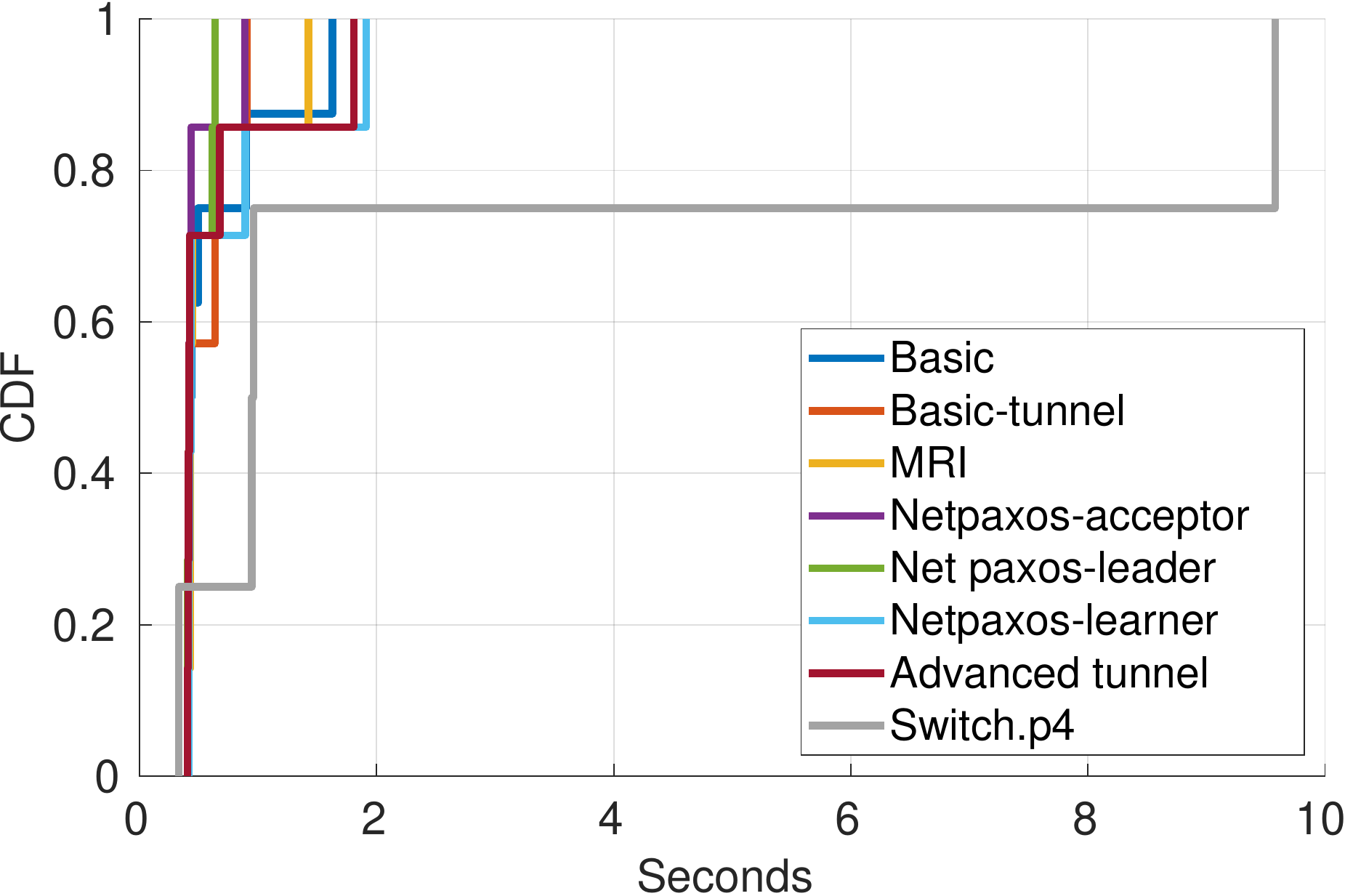}
\subcaption{Median of bug detection times (bmv2). \label{fig:med_det}}
\end{subfigure}
\hspace{0.01\textwidth}
\begin{subfigure}[r]{0.66\columnwidth}
\centering
\includegraphics[width=\columnwidth]{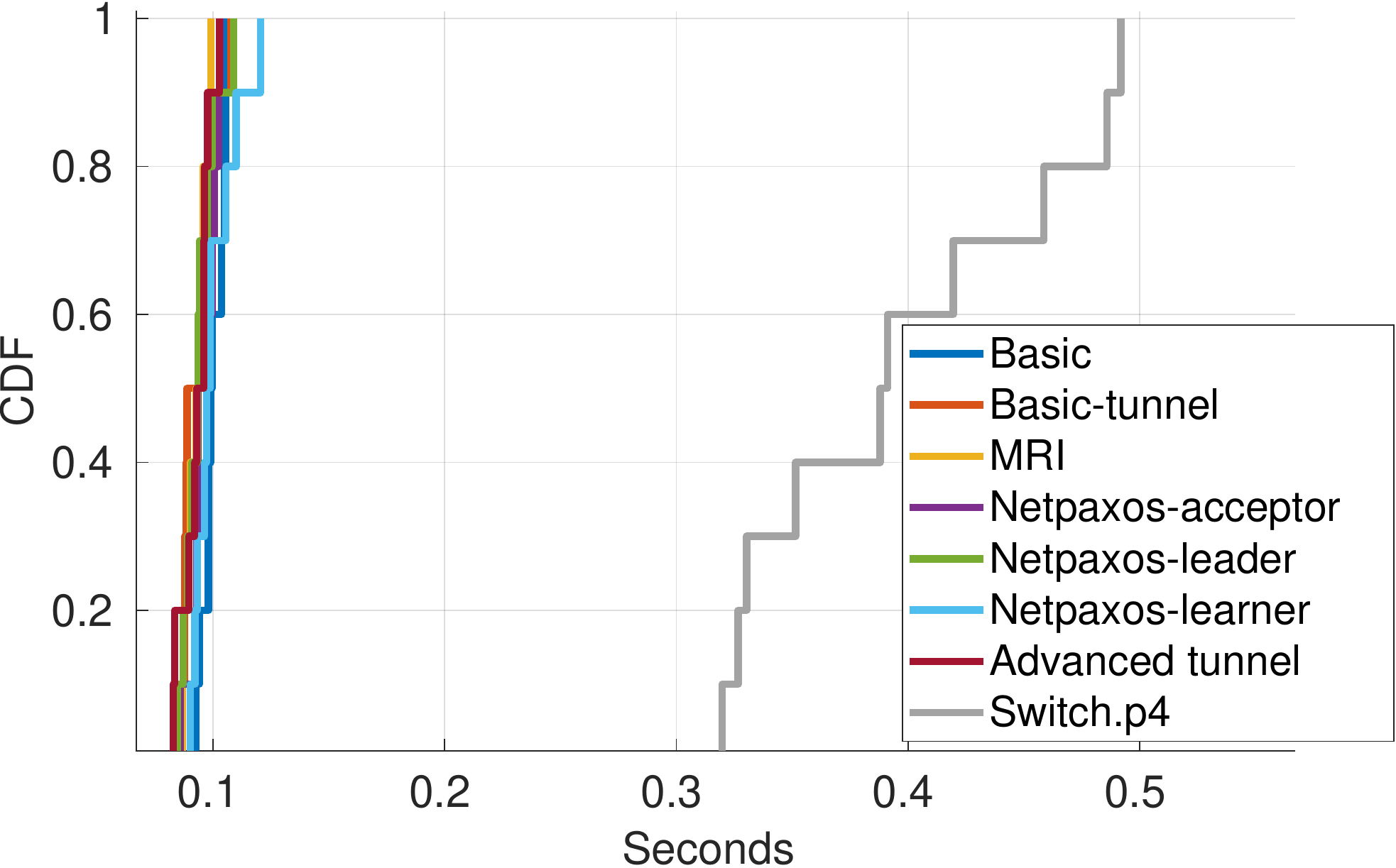}
\subcaption{Median of bug localization times (bmv2).\label{fig:med_loc}}
\end{subfigure}
\hspace{0.01\textwidth}
\begin{subfigure}[l]{0.66\columnwidth}
\centering
\includegraphics[width=\columnwidth]{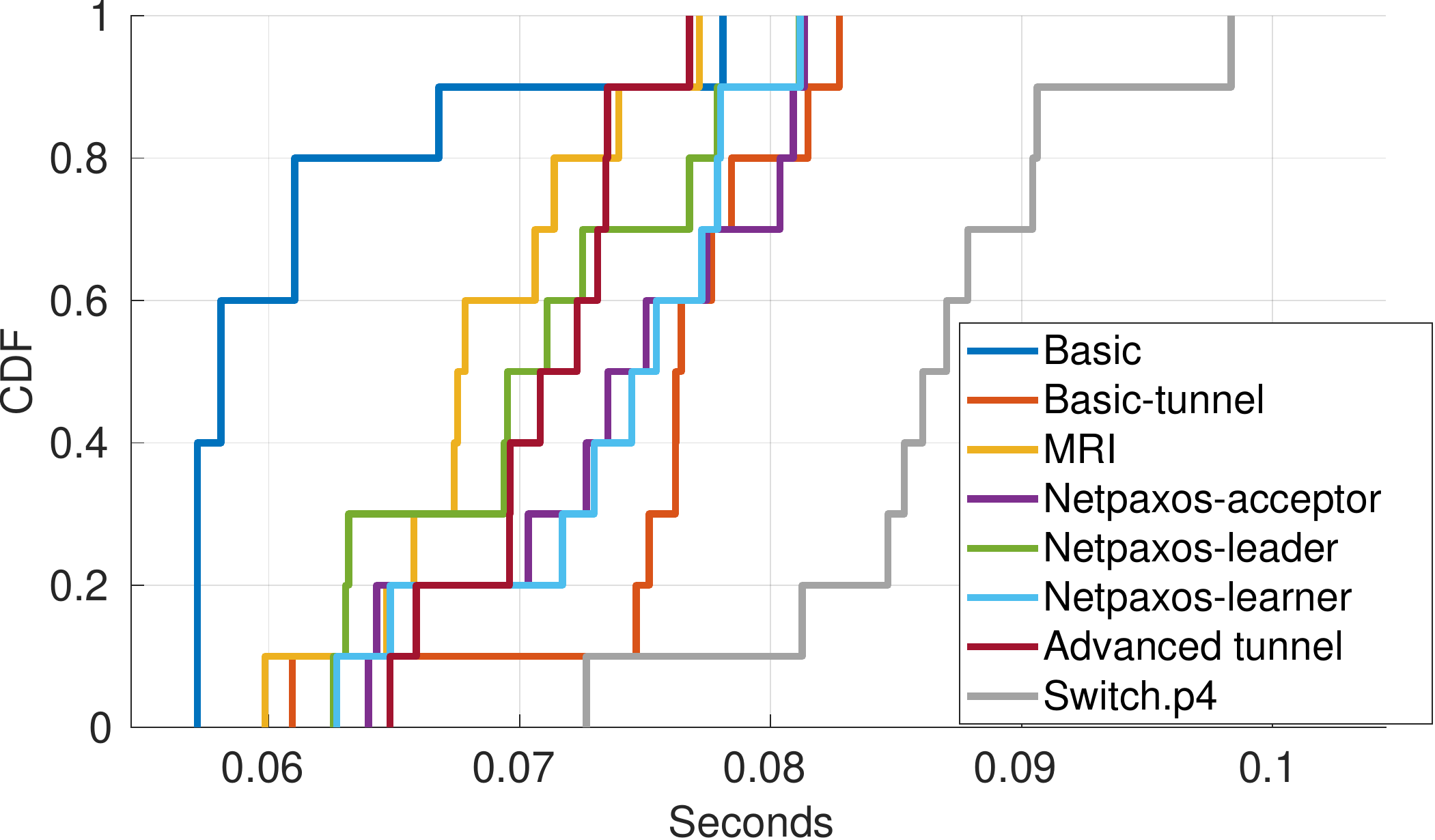}
\subcaption{Median time of patching the code (bmv2).\label{fig:med_patch}}
\end{subfigure}
\centering
\begin{subfigure}[l]{0.66\columnwidth}
\centering
\includegraphics[width=\columnwidth]{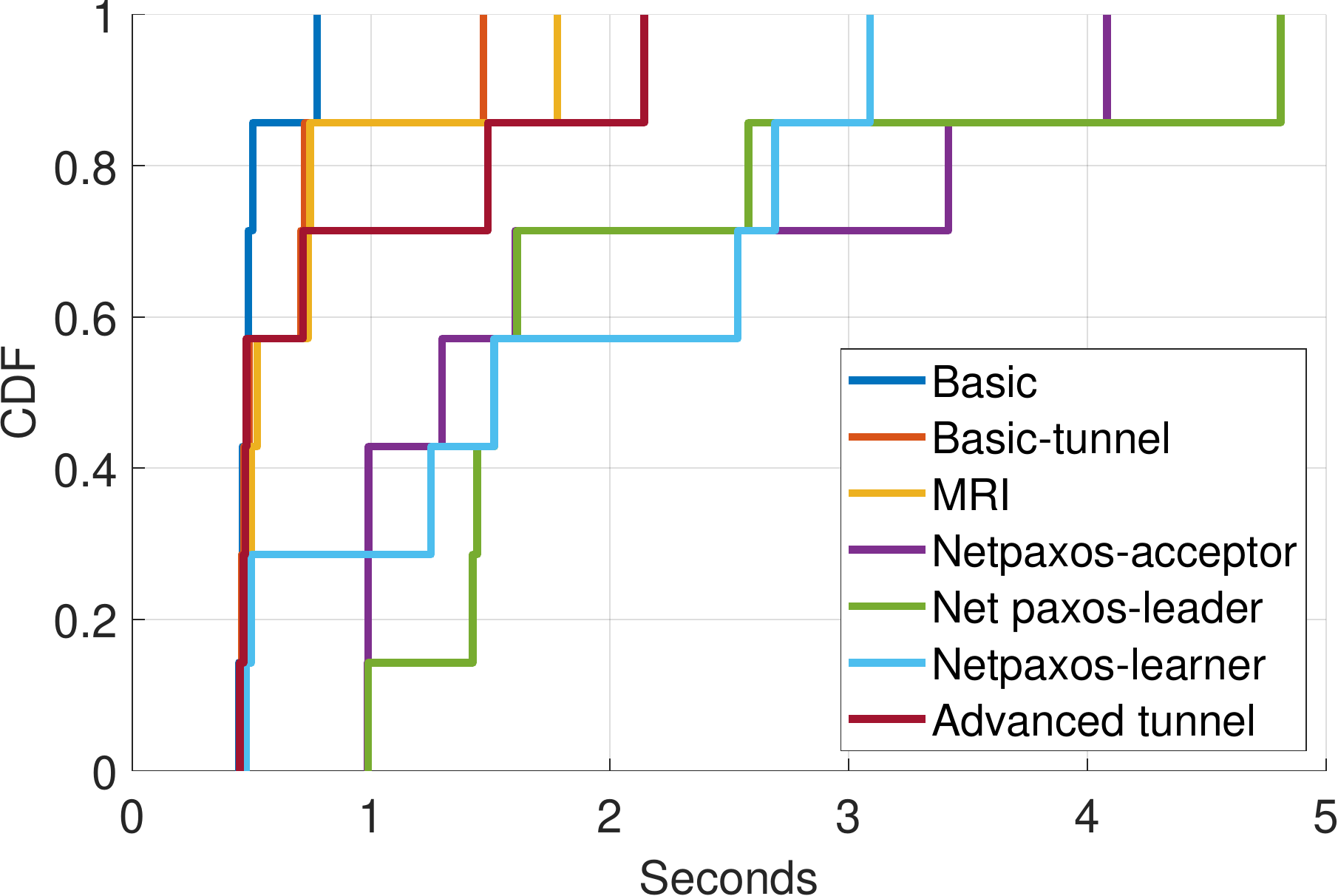}
\subcaption{Median of bug detection times (Tofino). \label{fig:med_det2}}
\end{subfigure}%
\hspace{0.01\textwidth}
\begin{subfigure}[c]{0.66\columnwidth}
\centering
\includegraphics[width=\columnwidth]{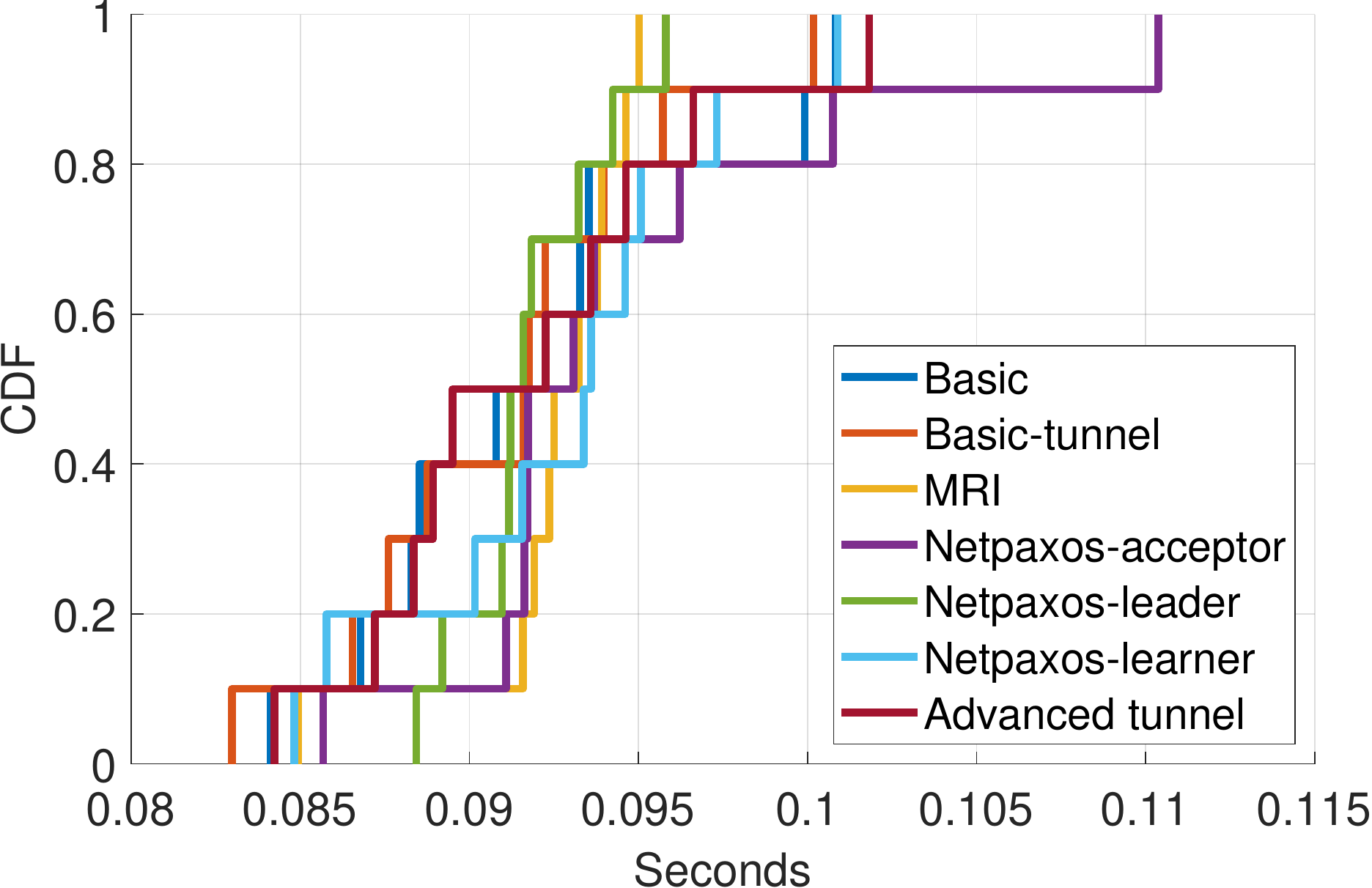}
\subcaption{Median of bug localization times (Tofino).\label{fig:med_loc2}}
\end{subfigure}%
\hspace{0.01\textwidth}
\begin{subfigure}[r]{0.66\columnwidth}
\centering
\includegraphics[width=\columnwidth]{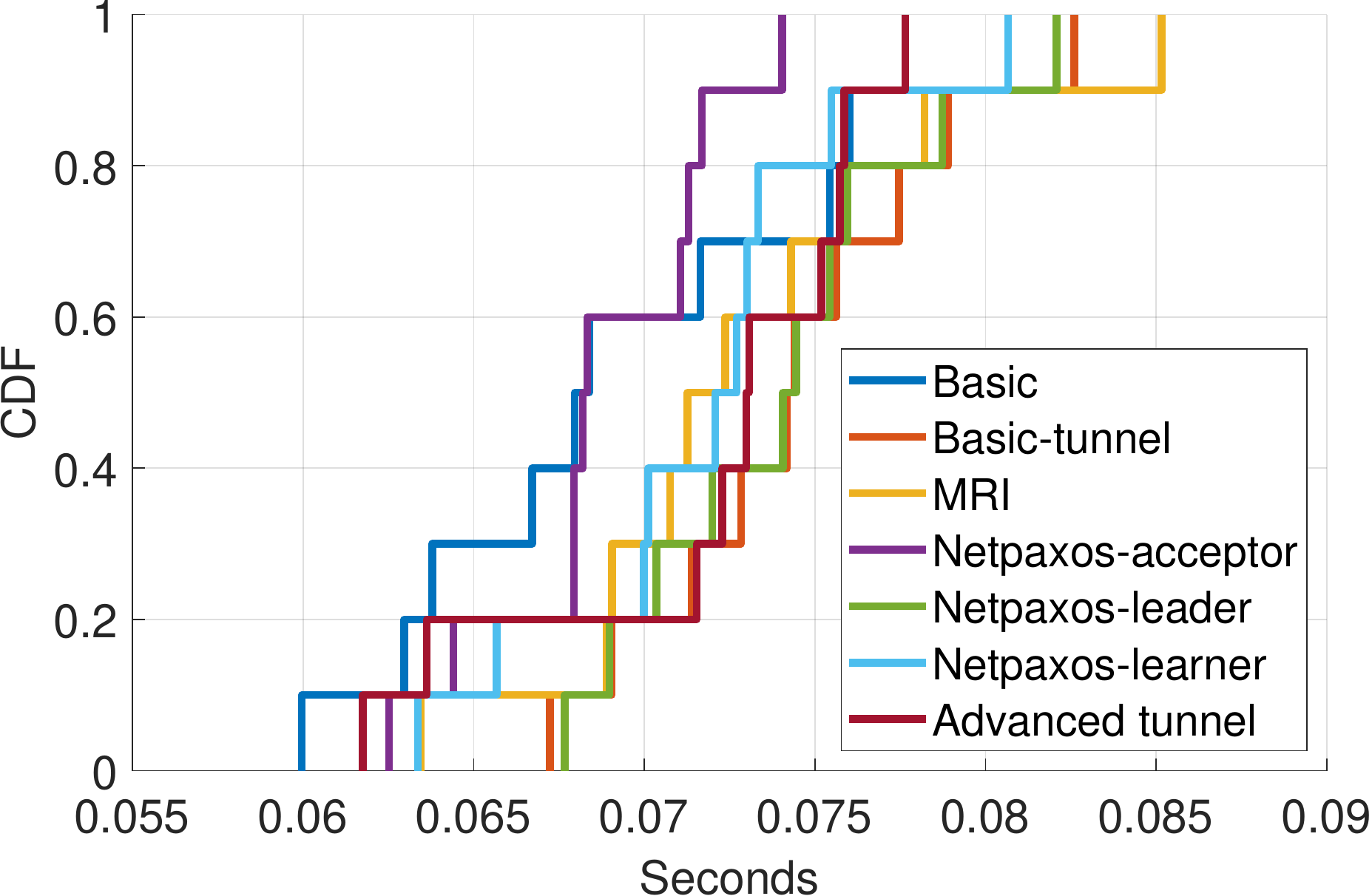}
\subcaption{Median time of patching the code (Tofino).\label{fig:med_patch2}}
\end{subfigure}
\hspace{0.01\textwidth}
\caption{Bug detection, localization and patching times of different P4 programs in bmv2 and Tofino. Each plot represents a median over 10 runs.}
\label{fig:evaluation_results}
\end{figure*}
\begin{figure*}[t]
\centering
\begin{subfigure}[r]{0.66\columnwidth}
\centering
\includegraphics[width=\columnwidth]{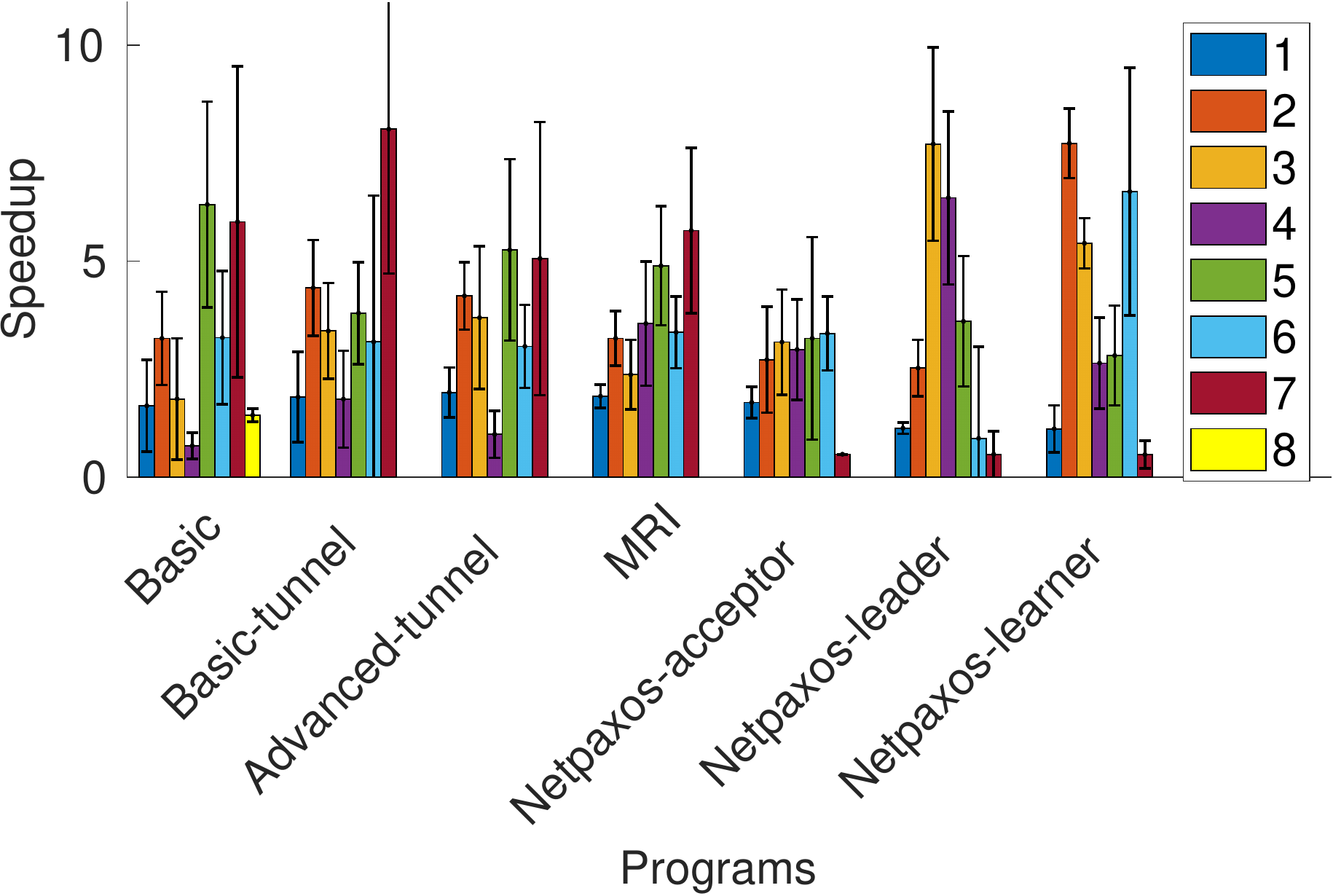}
\subcaption{Speedup of median bug detection times: \system vs Advanced Agent. Bug IDs (Table~\ref{tab:packet_types}) in the legend. \label{fig:actual_speedup_simple}}
\end{subfigure}
\hspace{0.01\textwidth}
\begin{subfigure}[r]{0.66\columnwidth}
\centering
\includegraphics[width=\columnwidth]{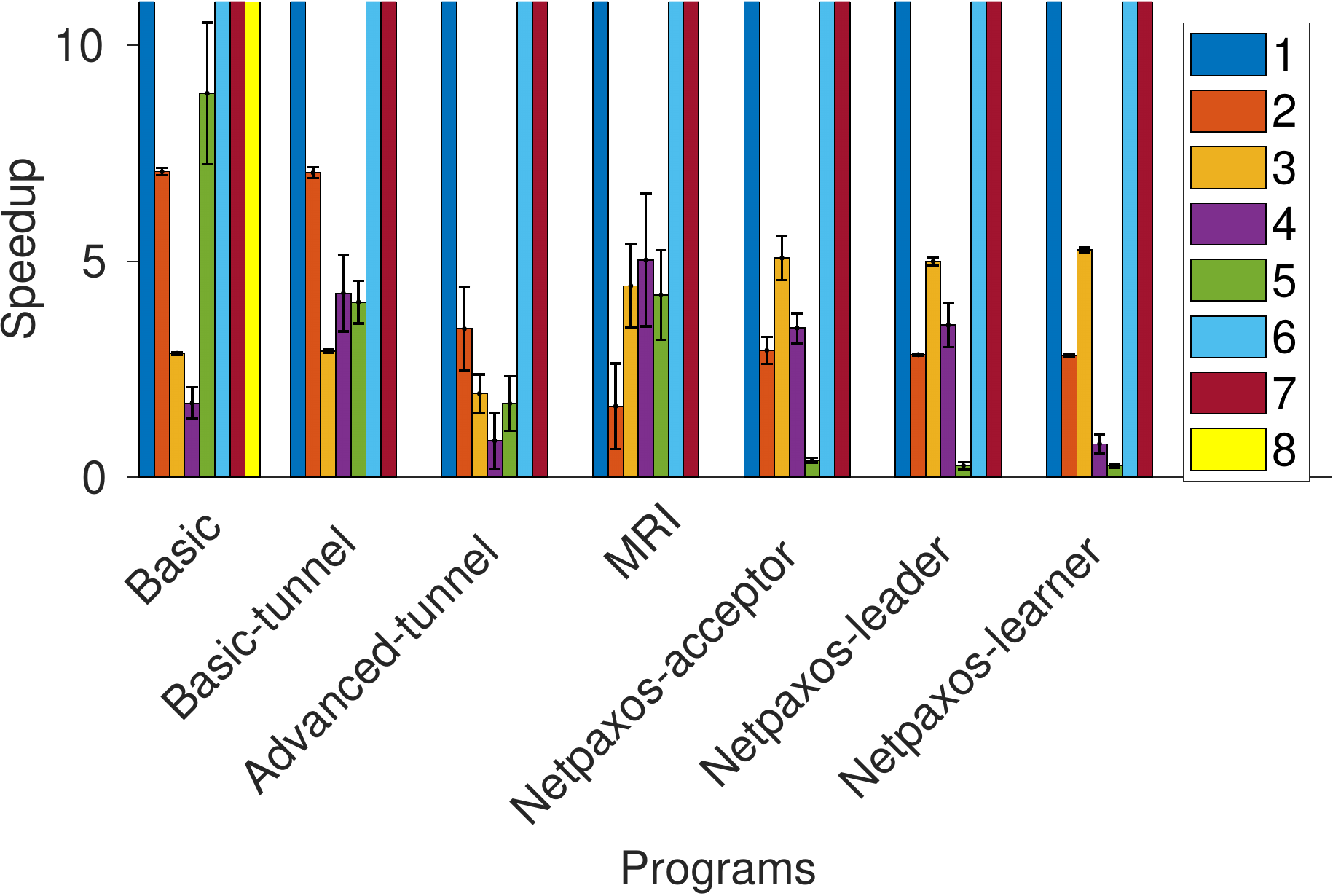}
\subcaption{Speedup of median bug detection times: \system vs IPv4-based fuzzer. Bug IDs (Table~\ref{tab:packet_types}) in the legend.\label{fig:actual_speedup_IPv4}}
\end{subfigure}
\begin{subfigure}[r]{0.66\columnwidth}
\centering
\includegraphics[width=\columnwidth]{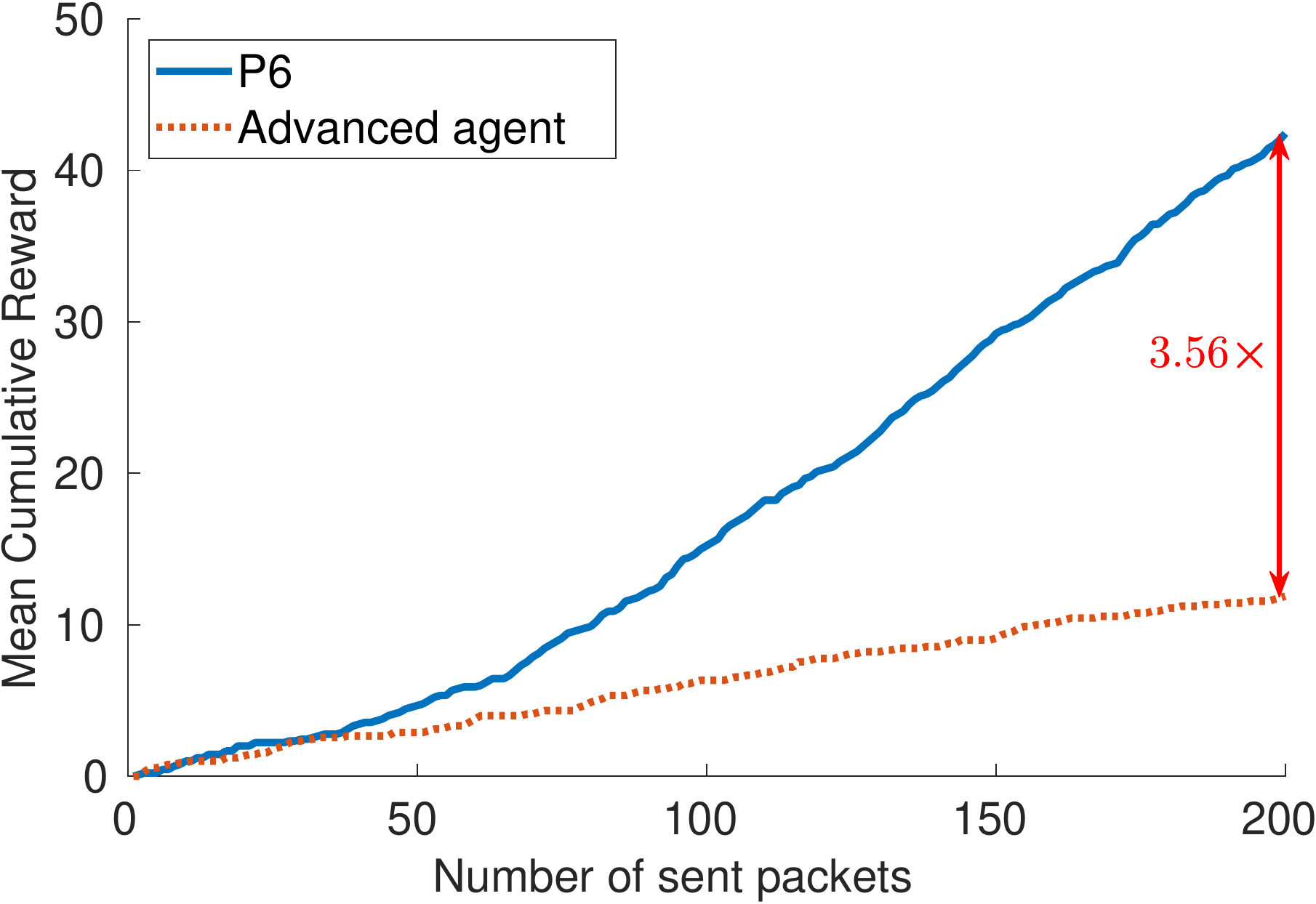}
\subcaption{Training: \system vs Advanced Agent (random action selection) in the case of bug 4 (Table~\ref{tab:packet_types}). \label{fig:rewards_training}}
\end{subfigure}
\caption{\system vs Baselines. Each plot represents a median over 10 runs.}
\label{fig:speedup}
\end{figure*}
\begin{table}
\centering
\scriptsize
\scalebox{0.95}{
\begin{tabular}{|*{5}{c|}}
\hline
\new Applications & bmv2 (LOC) & Tofino (LOC) & $\%$ increment\\
\hline
\tt{basic.p4}~\cite{p4tut} & $181$ & $257$ & $41.9$  \\
\hline
\tt{basic\_tunnel.p4}~\cite{p4tut} & $219$ & $302$ & $37.9$ \\
\hline
\tt{advanced\_tunnel.p4}~\cite{p4tut} & $242$ & $316$ & $30.6$ \\
\hline
\tt{mri.p4}~\cite{p4tut} & $277$ & $372$ & $34.3$ \\
\hline
\tt{netpaxos-acceptor.p4}~\cite{netp} & $270$ & $327$ & $21.1$ \\
\hline
\tt{netpaxos-leader.p4}~\cite{netp} & $259$ & $323$ &  $24.7$\\
\hline
\tt{netpaxos-learner.p4}~\cite{netp} & $288$ & $355$ & $23.2$ \\
\hline
\tt{switch.p4}~\cite{switch} & $8715$ & -\tablefootnote{\texttt{switch.p4} is not publicly available for Tofino.} & - \\
\hline

    \end{tabular}}
    \caption{\new Programs' LOC in bmv2 and Tofino.} \label{tab:loc}
\end{table}
\subsection{Experiment Strategy}\label{sec:exp}
\begin{figure}[t]
\centering
\includegraphics[width=\columnwidth]{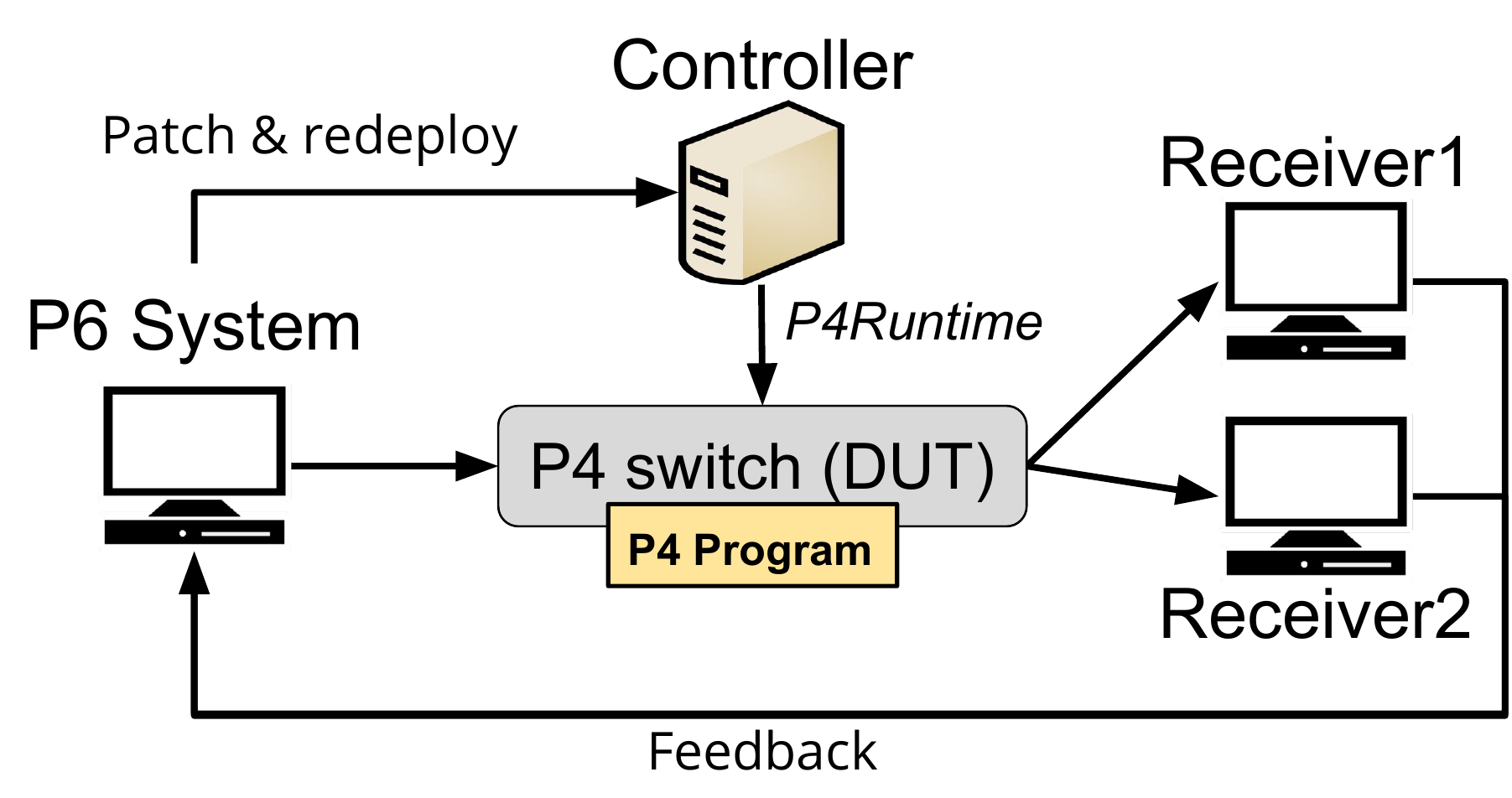}
\caption{Experiment Topology}
\vspace{1em}
\label{fig:prot}
\end{figure}
For conducting our experiments and to evaluate \system prototype, we ran \system together with the P4 switch and control plane module in a Vagrant~\cite{vagrant} environment with VirtualBox~\cite{virtualbox}. We emulate the network shown in Figure~\ref{fig:prot}. For each program, separate Vagrant machines, each with $10$ CPU cores and $7.5$ GiB RAM, are used. The Vagrant machines ran on a server running Debian 9 OS (Version 4.9.110), with Intel Xeon CPU and $256$ GiB RAM. Each experiment was executed ten times on each of the eight \new programs shown in Table~\ref{tab:loc} from P4 tutorials~\cite{p4tut}, NetPaxos~\cite{netp} and switch.p4~\cite{switch} repository. For each of the ten runs, $9$ test-cases were executed, where each test-case corresponds to one condition of the queries $1-6$ illustrated in Figure~\ref{fig:langf}. Note, \texttt{basic.p4} program has $10$ test-cases as it is also tested using query $7$ for the \pd bug. Furthermore, we currently trained the \agent separately for each test-case and sequentially execute the test cases. This can, however, be parallelized easily. We observed that for only two conditions of the \lang queries, no bugs were detected. 

\subsubsection{Experiment Topology}\label{sec:exp2}

Figure~\ref{fig:prot} shows the topology used for the experiments as part of the \system system evaluation. In total, five virtual machines are used. One of the machines is running the P4 switch, which is connected to all other machines. The controller is connected to the P4 switch, in order to deploy the P4 program and fill the forwarding tables. The \system system is connected to one port of the switch for sending the packets generated by \fuzz. Two virtual machines act as the receiver for these packets. The feedback of the receivers is used by the \system system to evaluate the actual behavior of the P4 switch. In addition, the machine running \system is connected to the controller to trigger re-deployment of the patched P4 program, in case a bug is detected.
\subsection{Metrics}\label{sec:metrics}
In particular, we ask the following questions:\\
\emph{\textbf{Q1.} How much time does} \system \emph{take to detect, localize, and patch all bugs?} (\cref{sec:perf1})\\
\emph{\textbf{Q2.} How does} \system \emph{perform against the baselines?} (\cref{sec:perf2})\\
\emph{\textbf{Q3.} How many rewards does} \system \emph{generate against the baseline of an Advanced Agent 
for 
\agent training?} (\cref{sec:perf3})\\
\emph{\textbf{Q4.} How many packets does} \system \emph{generate to detect bugs against the baselines?} (\cref{sec:perf4})\\ 
\emph{\textbf{Q5.} What is the accuracy of} \system\emph{?} (\cref{sec:perf5})
\subsubsection{Performance of \system}\label{sec:perf1}
To evaluate the performance of \system, we execute the detection, localization, and patching on $8$ publicly available P4 programs from the P4 tutorials~\cite{p4tut}, NetPaxos~\cite{netp} and switch.p4~\cite{switch} repository with minimal manual efforts.

Figure~\ref{fig:med_det} and \ref{fig:med_det2} show the median bug detection time of \system over ten runs for the different programs using bmv2 SimpleSwitchGrpc and Barefoot Tofino Model, respectively. Note, \texttt{switch.p4} program is only available for bmv2 and was not tested using Tofino.  In all runs on bmv2 except for \texttt{switch.p4} program, \system was able to detect all bugs in less than two seconds. In \texttt{switch.p4}, \system was able to detect all bugs in less than ten seconds. The detection time is higher for \texttt{switch.p4} as compared to the other tested programs since more packets get dropped making bug detection more difficult. On Tofino, the median detection time was slightly higher for four out of seven programs. The reason for the increased bug detection time with the NetPaxos programs~\cite{netp} can be due to the instrumentation of these programs by us to make them run on CPU intensive Tofino Model. 

Figures \ref{fig:med_loc} and \ref{fig:med_loc2} illustrate the median bug localization time of \system for the different programs using bmv2 SimpleSwitchGrpc and Barefoot Tofino Model. 
Overall, all bugs for $7$ of the programs were localized by \system in just above $0.12$ seconds on bmv2 and Tofino. To our surprise, the bug localization time for \texttt{switch.p4} program running on bmv2 is only increased by a factor of $4\times$, even though the program has about $30\times$ more lines of code compared to the other tested programs (see Table~\ref{tab:loc}).
The median time of patching the code is shown in Figures \ref{fig:med_patch} and \ref{fig:med_patch2} for bmv2 and Tofino respectively. \system is able to patch the P4 programs with millisecond scale performance (max. $98$ milliseconds). 
\subsubsection{\system vs Baselines: Detection Time}\label{sec:perf2}
We compare \system against the three baseline approaches in terms of bug detection time. We observe that the Advanced Agent baseline, see Figure~\ref{fig:actual_speedup_simple} (with quartiles), was able to detect all the bugs present in the tested programs, which is due to the similarity with the \system \agent. Advanced Agent, however, cannot learn from the rewards, hence generates more packets and thus, takes more time to detect the bugs than \system \agent. IPv4-based fuzzer was only able to detect $4$ out of $10$ bugs in the seven programs from~\cite{p4tut, netp}. For \texttt{switch.p4} program~\cite{switch}, IPv4-based fuzzer was able to detect $3$ out of $4$ bugs which were IPv4-based. In Figure~\ref{fig:actual_speedup_IPv4} (with quartiles), the speedup is defined as infinite for the test-cases where IPv4-based fuzzer could not detect the bug. Accordingly, the bars representing these test-cases range until the top of the figure.
\emph{Note, Na\"ive fuzzer was not able to detect bugs at all, even though generating 16k packets.}

Figure~\ref{fig:actual_speedup_simple} shows the speedup (Advanced Agent/\system \agent) for all bugs detected in the seven tested programs from~\cite{p4tut,netp}. The results show that \system \agent can detect bugs up to $10.96\times$ faster than the Advanced Agent baseline. Only bug $7$ was detected faster by the Advanced Agent in $3$ of the $7$ \new applications tested as the Advanced Agent needs less time for random action selection than \system \agent for intelligent action selection, based on its neural networks. In addition, Advanced Agent can make use of the same mutation actions and the pre-generated dict, hence when triggering the bug, 
the overall execution time will be slightly lower than that of \system \agent. In $94\%$ of the test-cases, Advanced Agent required more time and 
packets to detect the bugs than the \system \agent. For \texttt{switch.p4} program~\cite{switch}, the results show that \system \agent is able to detect bugs up to $30\times$ faster than the Advanced Agent baseline.

Figure~\ref{fig:actual_speedup_IPv4} shows the speedup (IPv4-based fuzzer/\system \agent) for all bugs detected in the seven tested programs from~\cite{p4tut,netp}. For the test-cases where IPv4-based fuzzer was able to detect the bug, we observe that in $89\%$ of the test-cases \system \agent is able to detect the bugs faster while sending significantly fewer packets. \system \agent outperforms IPv4-based fuzzer by up to $8.88\times$ even though IPv4-based fuzzer sends packets at a higher rate. For \texttt{switch.p4} program, the results show that \system \agent is able to detect bugs up to $30\times$ faster than IPv4-based fuzzer. 

\begin{table}
\centering
\scriptsize
\scalebox{0.95}{
\begin{tabular}{|*{5}{c|}}
\hline
\new Applications & \cellcolor[gray]{0.9}\system & Advanced Agent & IPv4-based & Na\"ive\\
\hline
\tt{basic.p4}~\cite{p4tut}   &\cellcolor[gray]{0.9}$13$ & $59$ & $8,035$ & $16,000$  \\
\hline
\tt{basic\_tunnel.p4}~\cite{p4tut} & \cellcolor[gray]{0.9}$11$ & $59$ & $6,044$ & $14,000$  \\
\hline
\tt{advanced\_tunnel.p4}~\cite{p4tut}  & \cellcolor[gray]{0.9}$12$ & $57$ & $6,038$ & $14,000$  \\
\hline
\tt{mri.p4}~\cite{p4tut} & \cellcolor[gray]{0.9}$10$ & $61$ & $6,058$ & $14,000$  \\
\hline
\tt{netpaxos-acceptor.p4}~\cite{netp} & \cellcolor[gray]{0.9}$11$ & $52$ & $6,021$ & $14,000$  \\
\hline
\tt{netpaxos-leader.p4}~\cite{netp} & \cellcolor[gray]{0.9}$9$ & $49$ & $6,024$ & $14,000$  \\
\hline
\tt{netpaxos-learner.p4}~\cite{netp} & \cellcolor[gray]{0.9}$12$ & $44$ & $6,026$ & $14,000$  \\
\hline
\tt{switch.p4}~\cite{switch} & \cellcolor[gray]{0.9}$28$ & $113$ & $2,132$ & $14,000$  \\
\hline
    \end{tabular}}
    \caption{\system vs Baselines. Median \#packets sent per run over 10 runs.} 
    \label{tab:packets_sent}
\end{table}

\subsubsection{\system vs Advanced Agent Training}\label{sec:perf3}
To verify that \system \agent is able to effectively learn to detect bugs, we compare \system \agent against an Advanced Agent, that only relies on random action selection. This makes Advanced Agent similar, but not as intelligent as \system \agent. Advanced Agent can still execute the same mutation actions but is not able to reason about which actions lead to maximized rewards. 
Figure~\ref{fig:rewards_training} 
shows a comparison of the \textit{mean cumulative reward} (MCR) of the training process of both agents for bug ID $4$ of Table~\ref{tab:packet_types}. We observe that the \system \agent is able to outperform the baseline by a factor of $3.56\times$ 
for the mentioned case. Especially, the \textit{prioritized experience replay} helps the \system \agent to quickly learn about which actions lead to reward, hence trigger bugs in the program. Since, the \system \agent is trained only using experiences which are \textit{valuable} for the training.

\system \agent is trained for each condition of each query described by Figure~\ref{fig:langf} using the same set of hyper-parameters. 

\begin{figure}[t]
\centering
\includegraphics[width=\columnwidth]{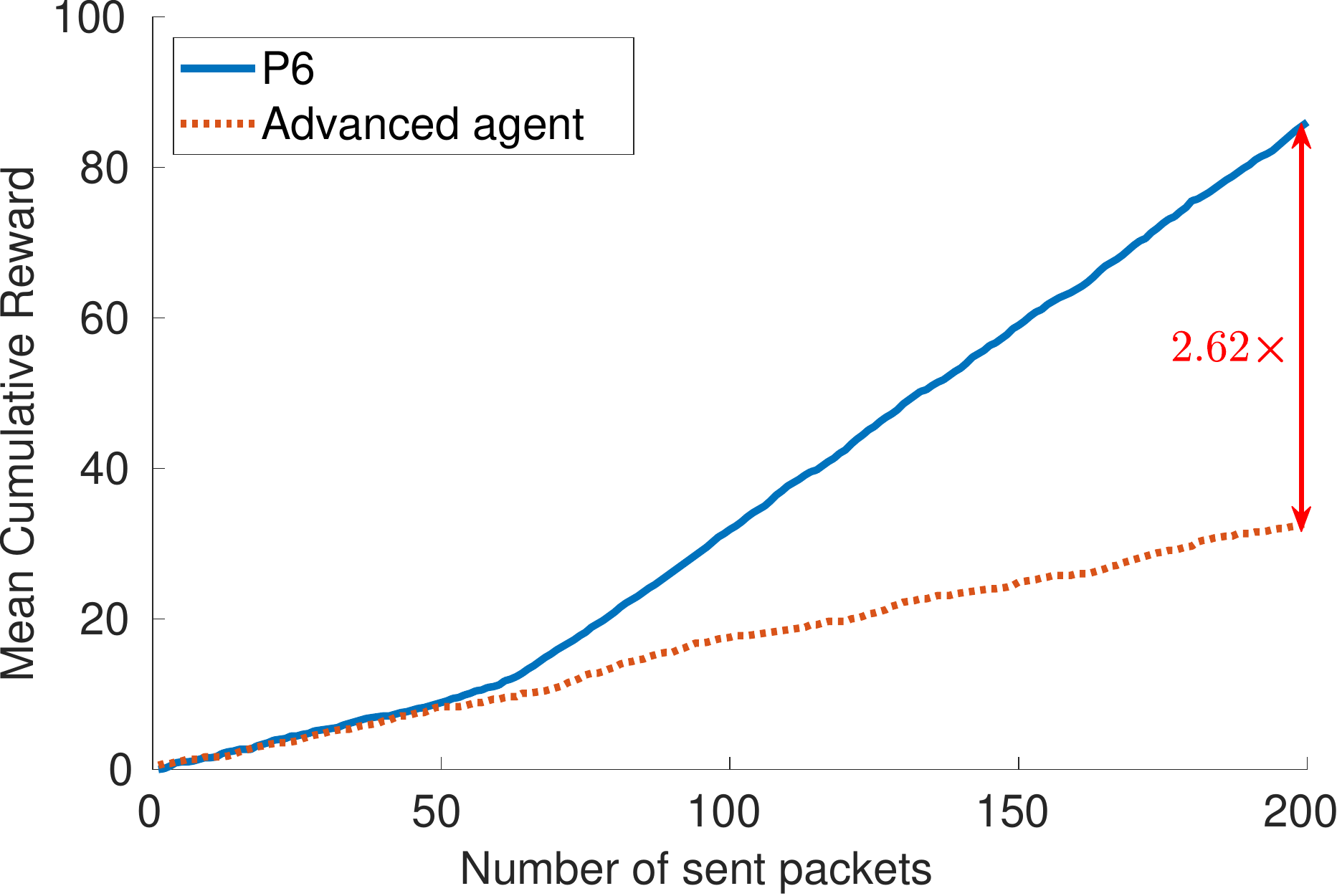}
\caption{Training: \system vs Advanced Agent (random action selection) in the case of bug ID 2 in Table~\ref{tab:packet_types}.\label{fig:rewards_training2}}
\end{figure}

\smartparagraph{More results.}
Figure~\ref{fig:rewards_training2} shows the training comparison results for the bug ID 2 (Generated wrong \texttt{checksum}) in Table~\ref{tab:packet_types}. Also, in this case, the \system \agent is able to outperform the Advanced Agent baseline by a factor of $2.62\times$. 

\begin{figure}[t]
	\centering
	\includegraphics[width=\columnwidth]{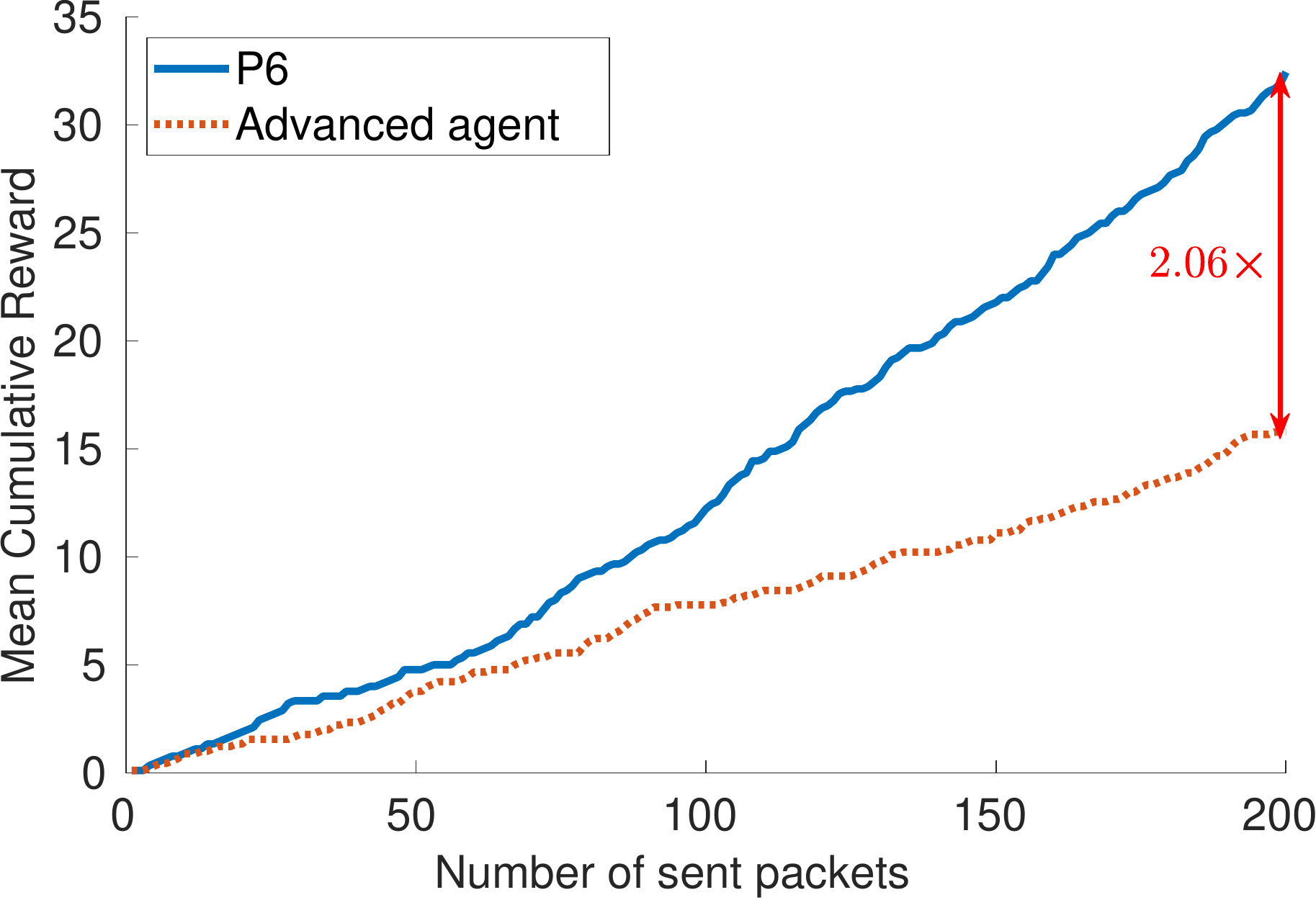}
	\caption{Training: \system vs Advanced Agent (random action selection) in the case of bug ID 5 in Table~\ref{tab:packet_types}.\label{fig:rewards_training5}}
\end{figure}

The training comparison results for the bug ID 5 (\texttt{IP TotalLen Value} is too small) can be seen in Figure~\ref{fig:rewards_training5}, showing that \system \agent is able to outperform Advanced Agent baseline by a factor of $2.06\times$.

\begin{figure}[t]
\centering
\includegraphics[width=\columnwidth]{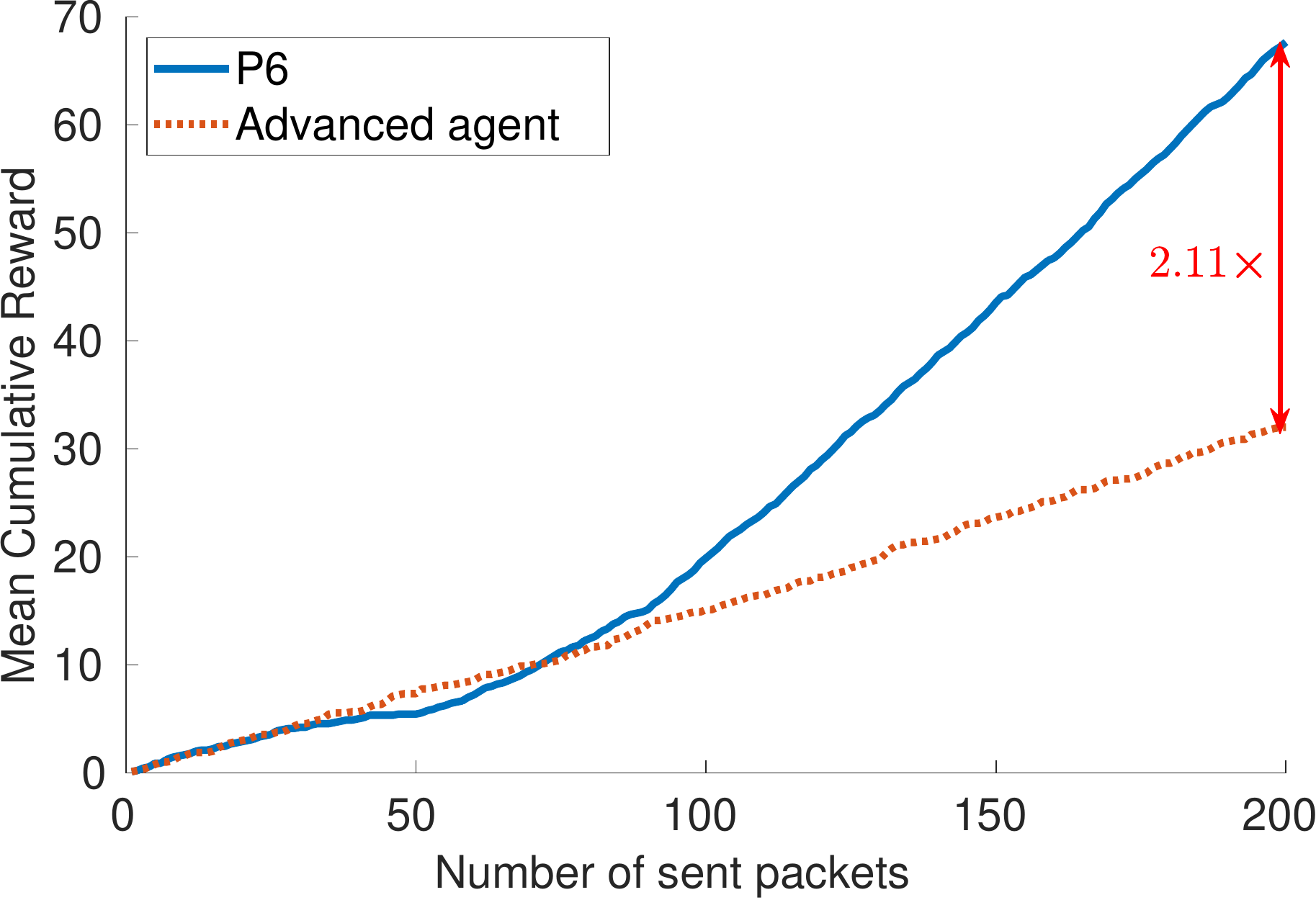}
\caption{Training: \system vs Advanced Agent (random action selection) in the case of bug ID 6 in Table~\ref{tab:packet_types}.\label{fig:rewards_training6}}
\end{figure}

In the case of the bug ID 6 (\texttt{TTL} 0 or 1 is accepted), \system \agent is able to outperform Advanced Agent baseline by a factor of $2.11\times$, as illustrated by Figure \ref{fig:rewards_training6}. \\

\begin{figure}[t]
\centering
\includegraphics[width=\columnwidth]{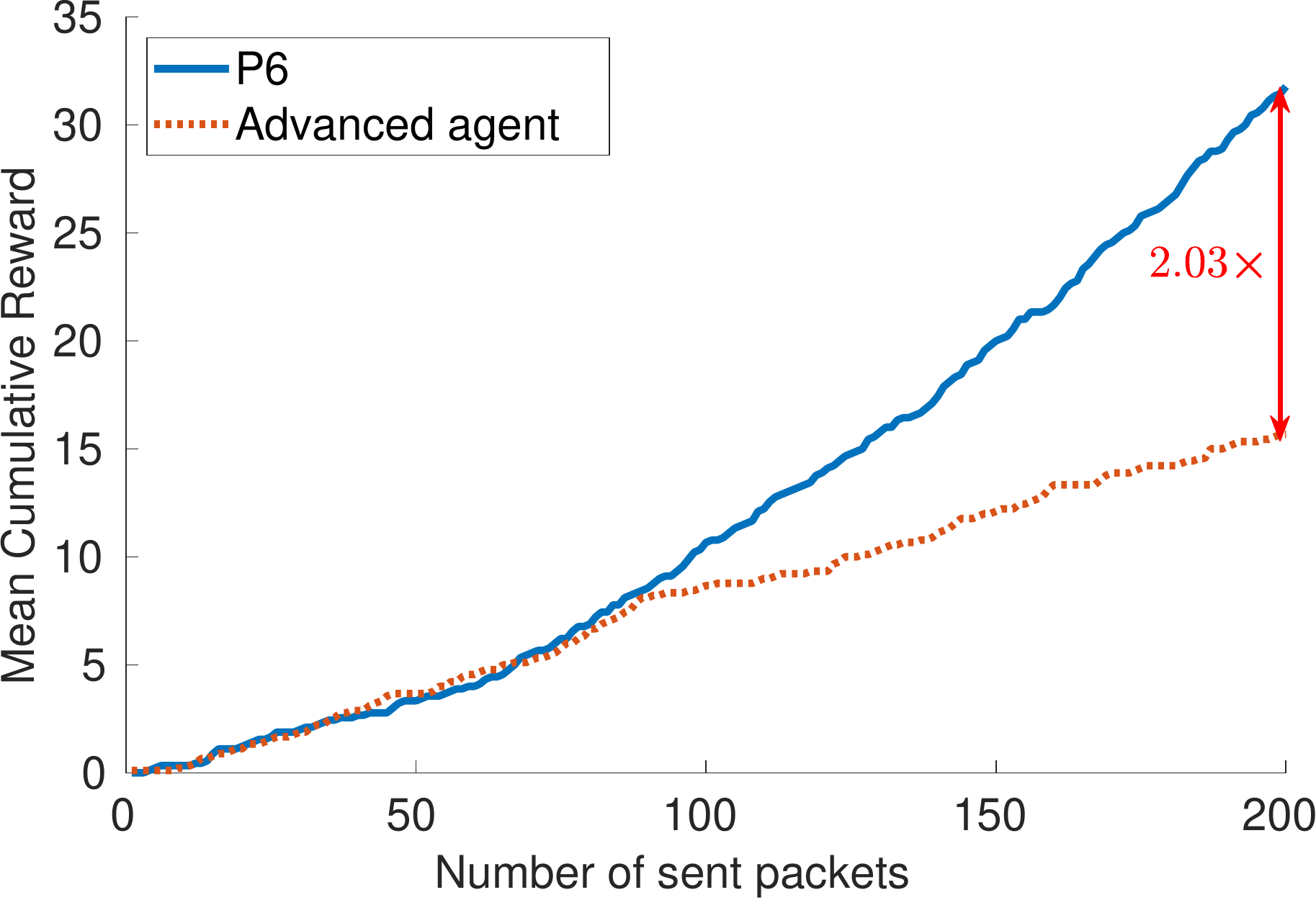}
\caption{Training: \system vs Advanced Agent (random action selection) in the case of bug ID 7 in Table~\ref{tab:packet_types}.\label{fig:rewards_training7}}
\end{figure}

Figure \ref{fig:rewards_training7} shows that \system \agent outperforms Advanced Agent for the bug ID 7 (\texttt{TTL} not decremented) by a factor of $2.03\times$. These results show the clear advantage of \system \agent over the Advanced Agent baseline. \\
Above results show \system \agent consistently outperforms Advanced Agent in MCR for other queries. 
Overall, this shows the clear advantage of \system \agent over the baselines and the ability to detect bugs with fewer packets. 
\subsubsection{\system vs Baselines: Dataplane Overhead}\label{sec:perf4}
Table~\ref{tab:packets_sent} illustrates the number of packets sent by \system and the baselines. This shows the usefulness of the \system \agent which generates less packets by learning about rewards, and generates packets that trigger bugs. In this case, the Advanced Agent is almost similar. IPv4-based fuzzer can detect $4$ out of $10$ bugs, but generates around $6k$ packets per run. For each test-case, na\"ive fuzzer sends around $2k$ packets (in total between $12k$ and $16k$) but it was not able to trigger any bug.
\subsubsection{\system Accuracy}\label{sec:perf5}
During the experiments, we could not observe any false positives in bug detection. Note, false positives can occur if the \lang queries are not correct and complete.
In the case of bug localization by \tar, false positives are not observed. 
We did not observe any false positives in patching by the \pat as it only fixes the code if the correct code is missing. Note, if localization results by \tar contain false positives, then \pat is prone to false positives as it makes the patching decisions based on the localization results.

\smartparagraph{Summary.}
Our results show that \system due to RL-guided fuzzing 
significantly outperforms the baselines across two different platforms: bmv2 and Tofino in terms of detecting runtime bugs (including \pd bugs) with minimal dataplane overhead non-intrusively. We observe that most of the \pin bugs existed in the parser or header part, otherwise packets with invalid headers get rejected. 
\system accurately and swiftly localizes and patches (millisecond scale) the 
bugs due to the P4 program structure in an automated fashion. 

\section{Related Work}\label{sec:rel}
Verification of programmable networks has been in a constant state of flux. Approaches like \cite{Kazemian2012,Kazemian2013,veriflow} perform modeling of the network from the control plane to check the reachability, loop freedom, and slice isolation. ATPG \cite{Zeng2014} generates test packets based on control plane configuration using~\cite{Kazemian2012} for functional and performance verification in traditional networks and SDNs (Software-defined Networks). All of the aforementioned tools~\cite{Kazemian2012,Kazemian2013,veriflow,Zeng2014}, however, assume that the control plane has a consistent or correct view of the data plane in traditional IP-based networks or SDNs only. \system does not assume the correctness of the control plane and observes the runtime behavior to detect, localize and patch the software bugs in P4 switches.~\cite{rajpal2017not,godefroid2017learn,cummins2018compiler} use different machine learning approaches for finding security vulnerabilities or compiler specific-bugs which cause crashes, however, they are insufficient for network-related verification. \system executes switch verification to identify the bugs in a P4 switch. 

Currently, 
most of the P4-related verification techniques, 
use static analysis of P4 programs using symbolic execution~\cite{stoenescu2018debugging,neves2018verification,Neves:2018:VPP:3281411.3281421} or Hoare logic~\cite{liu2018p4v}. The static analysis is prone to false positives as it analyzes the P4 program without passing any real inputs, e.g., packets. Therefore, checksum-related bugs where computations are required on input packets and \pd bugs cannot be detected. Such bugs require \system-like runtime verification. In addition,~\cite{liu2018p4v,neves2018verification,Neves:2018:VPP:3281411.3281421} require a P4 program to be manually annotated by the programmer which is cumbersome and prone to manual errors whereas \system is non-intrusive as it does not require to modify P4 program for bug detection and localization. p4pktgen~\cite{Notzli:2018:PAT:3185467.3185497} focuses on locating errors in the toolchains used to compile and run P4 code, e.g., \texttt{p4c}, and uses symbolic execution to create exemplary packets which can execute a selected path in the program. However, it cannot detect \pd bugs or egress pipeline bugs. Such a verification method can complement our solution. P4NOD~\cite{lopes2016automatically} statically models the network, however, it does not check how the actual P4 switches behave upon receiving the malformed packets e.g., incorrect IPv4 checksum. Cocoon~\cite{Ryzhyk:2017:CCN:3154630.3154686} suggests refinement-based programming for network verification. While this approach tries to ensure that programs match their specification, it requires a huge amount of additional and manual user input. For runtime verification, such a formal method is insufficient. Recently,~\cite{kodeswaran2020tracking} propose data-plane primitive for detecting and localizing bugs: tracking each packet’s execution path through the P4 program by augmenting P4 programs. However, this remains an in-progress and intrusive approach as it requires augmenting P4 code whereas \system does not change P4 program.
In-band network telemetry (INT)~\cite{kim2015band,int-specs} enables to collect telemetry data from each switch, however, unlike \system, it cannot localize or patch bugs if packets get dropped. Recently, Shukla et al. proposed P4CONSIST~\cite{p4consist}, a system that gathers the control and data plane states independently for comparison to verify the control-data plane consistency of P4 SDNs by detecting the path violations for critical flows in P4, however, without localization or patch support. 
In~\cite{shukla2}, a machine learning guided-fuzzing system is used to only detect \pin bugs in P4 programs. In the context of fuzzing, two approaches like~\cite{afl} and~\cite{tfuzz} are worth-considering. \cite{afl} is insufficient as it uses program coverage feedback to guide fuzzing without knowing which mutations lead to bugs. \cite{tfuzz} transforms the target program to remove sanity checks for fuzzing, however, it is intrusive as the target program requires modification for testing. 

Unlike \system, P4-based verification approaches~\cite{stoenescu2018debugging,liu2018p4v,neves2018verification,Neves:2018:VPP:3281411.3281421,Notzli:2018:PAT:3185467.3185497,lopes2016automatically,Ryzhyk:2017:CCN:3154630.3154686,kim2015band,int-specs,shukla2,p4consist} are insufficient in localizing and patching the runtime bugs in P4 programs. 
Besides, they cannot detect the \pd bugs. 
Table~\ref{table:related_work_summary} illustrates the capabilities of other P4 verification tools as compared to \system. 

\begin{table}[t!]
\centering
{\LARGE
		\scalebox{0.34}{
 \begin{tabular}{|*{7}{c|}} 
 \hline
 Related work in P4 & Runtime Verification & Detection & Localization & Patching & Detection of PD bugs \\ 
 \hline
 Cocoon~\cite{Ryzhyk:2017:CCN:3154630.3154686} & $\times$ & \checkmark & \checkmark & $\times$  & $\times$\\ 
 \hline
 Vera~\cite{stoenescu2018debugging} & $\times$ & \checkmark & $\times$ & $\times$ & $\times$\\ 
 \hline
 p4v~\cite{liu2018p4v} & $\times$ & \checkmark & $\times$ & $\times$ & $\times$\\ 
 \hline
 P4-ASSERT~\cite{Neves:2018:VPP:3281411.3281421,neves2018verification} & $\times$ & \checkmark & $\times$  & $\times$ & $\times$ \\ 
 \hline
 P4NOD~\cite{lopes2016automatically} & $\times$ & \checkmark & $\times$ & $\times$ & $\times$\\ 
 \hline
 p4pktgen~\cite{Notzli:2018:PAT:3185467.3185497} & $\times$ & \checkmark & $\times$ & $\times$ & $\times$\\ 
 \hline
P4CONSIST~\cite{p4consist} & \checkmark & \checkmark & $\times$ & $\times$ & $\times$\\ 
\hline
  \rl~\cite{shukla2} & \checkmark & \checkmark & $\times$ & $\times$ & $\times$\\ 
 \hline
\cellcolor[gray]{0.9}\system & \cellcolor[gray]{0.9}\checkmark & \cellcolor[gray]{0.9}\checkmark & \cellcolor[gray]{0.9}\checkmark & \cellcolor[gray]{0.9}\checkmark & \cellcolor[gray]{0.9}\checkmark\\ 
 \hline
\end{tabular}}
\caption{Related work in P4 verification. PD corresponds to the \pd bugs. Note, \checkmark denotes the capability, 
and $\times$ denotes the missing capability.}
\label{table:related_work_summary}}
\end{table}


\vspace{-.65em}
\section{Discussion}\label{sec:discussion}
\vspace{-.5em}
Traditionally, fuzz testing or fuzzing is known to offer a partial testing solution as it is prone to false negatives. Rice's theorem~\cite{hopcroft2006automata} states that all the non-trivial, semantic properties of a program are undecidable. Semantic properties refer to the behavior of a target program for all inputs. Therefore, if fuzzing does not detect any problems, it does not ensure that there is no problem at all. Statistical techniques like Good-Turing frequency estimations~\cite{good1953population,gale1995good,bohme2019assurance} for fuzz testing partially, aid in inferring the probability that the next generated test input leads to the discovery
of a previously unseen species.

In general, the quality of the input seeds, e.g., the relevance of mutations to the target and program coverage serve as good indicators to assess the quality of a fuzzer. However, there is a tradeoff between speed and precision of fuzz testing as there is an instrumentation overhead involved in generating a dictionary of meaningful inputs for effective testing and significantly reducing the input search space as compared to random mutations of bits in the inputs.   

Machine learning techniques like reinforcement learning~\cite{sutton2018reinforcement,Russell:2009:AIM:1671238} helps to some extent with the training of the models based on the feedback from the target. However, \emph{what is a good feedback?} is debatable. Traditionally, feedback depends on the coverage but it all boils down to the program under test. In addition, machine learning is as good as the input training data and thus, offers insufficient guarantees as instead of learning, the model may memorize. 
A generalized model applicable to \emph{any} kind of training data is highly desirable as it avoids the 
problems of overfitting and underfitting~\cite{burnham2011aic}.

Dynamic program analysis technique for fault-localization like Tarantula~\cite{jones2001visualization, jones2002visualization,jones2005empirical} benefits from utilizing the information from multiple failed test
cases which helps it in leveraging the richer information
base. Therefore, it helps to have at least one failed test case. In addition, allowing tolerance for passed test cases that
occasionally execute faults is essential for an effective fault-localization technique.

Software patching facilitates in fixing the software code errors. The patches, however, \emph{may} cause regressions which reflect in the form of abnormal behavior of the software. Sanity testing and modular code design facilitate in ensuring that basic functionality is not affected by the patch. However, one cannot assure that there are no other problems (false negatives) caused by the fix. 

We note that leveraging programmability, future \sloppy{programmable} networks will encompass even more possibilities of faults with a mix of vendor-code, reusable libraries, and in-house code. As such the general problem of network verification will
persist and we will have to explore how to extend the \system system to traditional IP-based networks. 

\section{Conclusion}\label{sec:conclusion}
We presented \system, the first system that enables runtime verification of P4 switches in a non-intrusive fashion. \system uses static analysis- and machine learning-guided fuzzing to detect multiple runtime bugs which are then, localized and patched on the fly with minimal human effort. 
Through experiments on existing P4 application programs, we showed that \system significantly outperforms the baseline bug detection approaches to detect existing \pin and -dependent bugs. In the case of \pin bugs, \system takes advantage of the increased programmable blocks to localize them and repair the \new programs, if and when a patch is available. 

We believe \system is an important foray into self-driving networks~\cite{feamster2017and}, which come with stringent requirements on dependability and automation. With \system, developers of P4 programs and operators of P4-enabled devices can improve the security of their products. As a part of our future agenda, we plan to apply \system on commercial-grade P4 programs and networks to report on our experience. We will also release the \system software and library of ready patches that we, respectively, developed and used in this work.

\bibliographystyle{unsrt}
\bibliography{refs}

\end{document}